\documentclass[12pt]{article}

\usepackage{amsmath}
\usepackage{amssymb}
\usepackage{graphicx}
\usepackage{bm}
\usepackage{xcolor}
\usepackage{mathtools}

\usepackage{times}

\newenvironment{sciabstract}{%
\begin{quote} \bf}
{\end{quote}}

\title{\vspace{-42.5mm}First passage times of transport on planar spatial networks and their connections to off-network planar diffusion}

\author
{Daniel B. Wilson,$^{1,2,\ast,\dag}$ C. H. L. Beentjes,$^{1,\dag}$\\
\\
\normalsize{$^{1}$Wolfson Centre for Mathematical Biology, Mathematical Institute, University of Oxford, }\\
\normalsize{Radcliffe Observatory Quarter, Oxford OX2 6GG, United Kingdom.}\\
\normalsize{$^{2}$Boston University, Department of Mathematics and Statistics, }\\
\normalsize{Boston, MA 02215, United States of America.}\\
\\
\normalsize{$^\ast$To whom correspondence should be addressed; E-mail: dbwilson@bu.edu.}\\
\normalsize{$^\dag$ These authors contributed equally}
}

\date{}

\begin{document} 


\baselineskip24pt


\maketitle


\begin{sciabstract}
Consider a network embedded in the 2D plane, where a particle diffuses along the  edges of the network. It is clear that over short length scales a particle moves along a single edge and thus undergoes one-dimensional diffusion. However, on larger length scales it is no longer immediately clear how the transport will behave. One could intuit that as the network is embedded in two dimensions for "large enough" length scales the transport will also appear two-dimensional. Is this true  for all networks? Can we quantify the length scales upon which this transition occurs? What is the transport behaviour on intermediate spatial scales? In this paper, we answer these question by presenting a numerical linear algebra approach that provides the exact moments of first passage times for a given network. Comparing these networked first-passage times to first-passage times for planar diffusion reveals several interesting properties of networked transport. In particular we can directly quantify the length scale upon which networked diffusion will appear planar if it does at all. Finally, we introduce an adaptation of the method of maximum entropy to use the moments of first-passage times to construct an analytical approximation to the first-passage times entire probability distribution. 
\end{sciabstract}

\newpage

\section{Introduction}\label{sec:intro}


Transport of individuals within complex environments occurs across all spatial scales. Examples include public transport through urbanised cities \cite{Joviae_Tranpsort10}; sediment flow through the pore-space in naturally occuring rock formations \cite{Blunt2002_AWR}; ionic exchange in lithium ion batteries \cite{Batteries}; and the coordinated transport of nanoparticles within the internal geometry of a cell \cite{Nano}. For each of these examples networks have offered convenient quantitative characterisations of complex environments \cite{DongPRE09,WilsonCommPhys21}, through which to study emergent transport behaviour and reveal how these behaviours relate to the underlying complex spatial structures. Due to the wide-spread interest in transport processes within networked topologies, there has been a large research focus within the mathematical and physical sciences to understand the interplay between networked topologies and transport statistics of interest \cite{Masuda2017,Wilson2018_PRE_A,Wilson2019_SIAM_JAM}.


A conical statistic that characterises a transport process is the first passage time (FPT). This is the time taken for an individual to pass over a boundary within the complex environment or network for the first time. In biology FPTs have been used to quantify the time taken for biomoelcules to navigate the cellular environment and locate the nucleus \cite{Jingwei2020}, as well as the time taken for ligand-antigen binding interactions to occur on a cell's surface \cite{Newby_PRL16}, and the time taken for protein molecules to locate specific sites on DNA  \cite{Shin_JCP19}. The applicability of FPTs to studying such a variety of biological transport processes has resulted in several studies exploring FPTs on complex networks \cite{Hwang_PRL12,Lau_EPL10}. 




The FPT for an individual on a network is the time taken for an individual undergoing some transport process to hit a subset of the networks nodes that constitutes the  first-passage boundary, often referred to as the boundary or target nodes. The global mean FPT (GMFPT) is the average time taken for an individual to hit a single target node averaged over all possible initial conditions. As such the GMFPT does not include any information about the iniital position of an individual. Applications of statistical physics to GMFPTs for a single absorbing node in complex networks have revealed rigorous analytical bounds that the GMFPT must lie between \cite{Tejedor_PRE09}. Morevover, a closed form solution for the GMFPT for a discrete random walk on a Vicsek fractal has been obtained using eigenvalues of the Laplacian matrix \cite{Zhang_PRE10}. However, the analytical methods used to study GMFPTs are often incapable of accurately quantifying  the mean FPT (MFPT) of an individual when the initial position is known. A numerical method using psuedo-Green functions has been developed as a computational tool to estimate MFPTs of random walkers in networks as well as higher moments of the FPTs \cite{Condamin_PRL05}. Finally, closed form analytical expressions that approximate the entire first passage time distribution on networks have been found using Laplace transformations \cite{Ding18}. However, both the pseudo-Green function and the Laplace transform approach are designed for discrete random walks where an individual hops from one node directly to neighbouring nodes with equal probability after a single fixed unit of time. As such, this technique does not extend to the study of FPTs of transport processes on networks where the distance between neighbouring nodes can be arbitrary, and the time taken to move between nodes is randomly distributed. 


In this paper we introduce a numerical method to study the first passage times of  individuals that diffuse along the edges of networks embedded within a two-dimensional Euclidean plane (the techniques developed however do extend to higher dimensions). The spatial networks we consider are very general other than the one condition that they be planar, i.e. no two edges can intersect. What is critical for our methodology to be applicable to such a broad range of networks is understanding the first-passage properties of a diffusing particle on a star graph, which to the authors knowledge have not been calculated before in the literature. Our methodology is then used to compare the FPTs of diffusing particles on networks to the FPTs of particles diffusing off-network in the plane. Such comparissons allow for a macroscopic understanding of networked diffusion. 


The remainder of this paper is organised as follows. In Section~\ref{sec:spatial-networks} we present the different classes of spatially embedded networks that we consider throughout this paper. In Section~\ref{sec:mfpt}, to study first passage properties of networked transport we present a general set of algebraic equations that can be used to solve for the mean first passage time. For a diffusive random walker we calculated the MFPT on each class of network as a function of the spatial scale of the network. Our results show that the MFPT behaviour on different networks varies greatly. In order to understand these results and make comparissons to first passage times for planar diffusion we need to consider higher moments. In Section~\ref{sec:higher-moments} we present a method to calculate higher moments of general networked first passage times through a hierarchical system of linear algebraic equations. Through numerical calculation of the coefficient of variation for networked first passage times we can compare transport on networks to planar transport. For the majority of networks as the size of the network is increased there is a transitiion from effective one-dimensional planar diffusion to two-dimensional planar diffusion. Our approach allows us to quantify both an effective diffusion coefficient in the large size limit as well as a length scale upon which this transition in effective dimensionality of the transport occurs. Diffusion on tree networks do not transition to two-dimensional diffusive behaviour and effective diffusion coefficients decay on larger and larger length scales. Therefore in Section~\ref{sec:radial-diffusion} we extend our analysis to compare diffusive transport on trees to radially dependent planar diffusion (a diffusive process where the diffusion coefficient decays as a function of the distance to the origin). In Section~\ref{sec:skewness-kurtosis-distributions} we consider higher moments of networked first passage times and explore distribution reconstruction algorithms based upon maximising the entropy to produce analytical approximations to the entire first passage time distribution for networked transport. Finally, in Section~\ref{sec:discussion} we summarise our results and discuss the scope for future work.

\section{Spatial network sampling}\label{sec:spatial-networks}


\subsection*{Poisson point networks}

In this section we consider spatial networks that arise from a randomly distributed collection of nodes. Let the plane $\mathbb{R}^2$ be populated by an infinite ensemble of points with position vectors $\mathcal{X} = \left\lbrace \vec{x} \in \mathbb{R}^2 \right\rbrace$ sampled from a Poisson point process with intensity parameter $\lambda_{p}$. A Poisson process is defined such that for a finite domain $\Omega \subset \mathbb{R}^2$ the number of Poisson points that lie within $\Omega$ is given by a Poisson random variable with rate parameter $\lambda_{p} |\Omega|$, and the position vectors are sampled independently and uniformly at random within the domain. These points can be used to build a vast array of spatially embedded networks that we will refer to as \textit{Poisson point networks}.

\begin{figure}
\centering
$\begin{array}{c}
\includegraphics[width=\columnwidth]{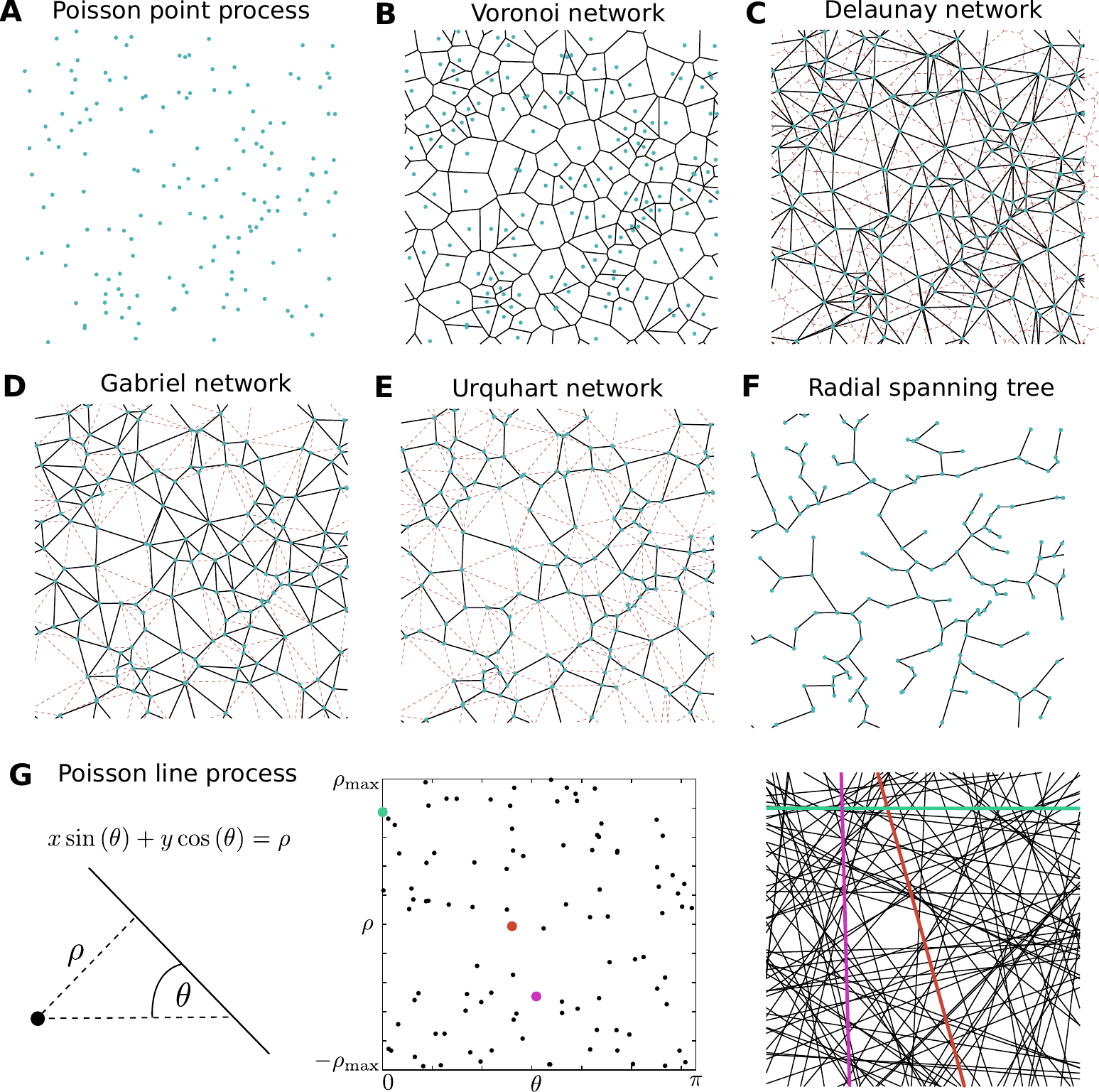}
\end{array}$
\caption{Stochastic planar network constructions in a two-dimensional plane. (A) A realisation of Poisson points (green dots). (B) A Voronoi network constructed from the realisation of Poisson points seen in panel A. (C) A Delaunay network constructed from the realisation of Poisson points seen in panel A, with the Voronoi network underneath (red dashed lines). (D) A Gabriel network constructed from the realisation of Poisson points seen in panel A, with the Delaunay network underneath (red dashed lines). (E) An Urquhart network constructed from the realisation of Poisson points seen in panel A, with the Delaunay network underneath (red dashed lines). (F) A Radial spanning tree constructed from the realisation of Poisson points seen in panel A. (G) A diagram explaining the construction of a Poisson line network.}
\label{fig:samplenetworks}
\end{figure}

The first network we consider is called a Voronoi network, and is built from the famous Voronoi tesselation which partitions $\mathbb{R}^2$ as follows. For each Poisson point $\vec{x}_i \in \mathcal{X}$ define a region $R_i = \left\lbrace \vec{x} \in \mathbb{R}^2 : ||\vec{x}-\vec{x}_i|| \leq ||\vec{x}-\vec{x}_j||, \forall j \neq i \right\rbrace$. Each region $R_i$ corresponds to the subset of $\mathbb{R}^2$ that is closer to the Poisson point $\vec{x}_i$ than any other Poisson point. Collectively these regions form polygons that partition the plane (see Fig.~\ref{fig:samplenetworks}B). The boundaries of these polygonal regions form the edges of the Voronoi network, and subsequently the nodes are placed where two distinct edges meet. The Voronoi tesselation has been used to describe the geometry formed by a colony of cells grown to affluence when initially seeded on a Petri dish \cite{Orozco}. Subsequently, studying transport processes along the Voronoi network is important when concerned with the transport of extracellular material such as solutes and nutrients that move in the empty space between cells.

Next, we consider the Delaunay network, otherwise known as a Delaunay triangulation. This network is the dual graph of the Voronoi network (see Fig.~\ref{fig:samplenetworks}C) and is constructed as follows. The Delaunay network takes the Poisson points as the networks nodes. Then two nodes with positions $\vec{x}_i$ and $\vec{x}_j$ are connected if the two Voronoi regions $R_i$ and $R_j$ share an edge, i.e. $R_i \cup R_j \neq \emptyset$. Similar to the Voronoi network, the Delaunay network describes the connectivity between adjacent cells. Thus, transport processes along the dual graphs (e.g. Delaunay networks) of tesselations (e.g. Voronoi networks) can be used to model resource sharing and communication between neighbouring cells \cite{Engblom2018_RSOS}.

In computational stochastic geometry there are countless other ways to connect Poisson points to form a spatially embedded network. Often these sampling methods will lead to networks that are subgraphs of the Delaunay network. We briefly consider three such networks. The Gabriel network is formed by connecting two Poisson points with positions $\vec{x}_i$ and $\vec{x}_j$ if the circle with diameter connecting these two Poisson points contains no other Poisson points. This network was first introduced to study geographic variational data \cite{GabrielPaper}, and is a subgraph of the Delaunay network (see Fig.~\ref{fig:samplenetworks}D). A further subgraph of the Gabriel network is the Urquhart network. This network is formed directly from a Delaunay traingulation where the longest edge in each triangle is removed (see Fig.~\ref{fig:samplenetworks}E). The Urquhart network was constructed purely to approximate the relative neighbourhood graph (which connects two Poisson points if and only if there is not a third Poisson point that is closer to both of them). However, the relative neighbourhood graph is notoriously difficult to construct, and the Urquhart network provides a computationally efficient approximation \cite{Andrade01}. 

The final network constructed from Poisson points that we consider is the radial spanning tree (RST) (see Fig.~\ref{fig:samplenetworks}F). To construct this tree, first a single Poisson point must be selected as the root of the tree, the plane is then translated such that this point lies on the origin for convenience. The remaining Poisson points are ordered by their Euclidean distance to the origin in increasing order. One-by-one the Poisson points in the ordered list are connected to the nearest Poisson point (in the Euclidean sense) that has already appeared in the list, i.e. is closer to the origin. If there are several Poisson points that satisfy this condition then one is selected uniformly at random.
\subsubsection*{Poisson line networks}


Here we consider an alternative construction of spatially embedded networks that focusses on sampling Poisson lines in $\mathbb{R}^2$ in lieu of Poisson points. Firstly we consider the parametrisation of a linear line in $\mathbb{R}^2$ by the angle $\theta$ to the $x$-axis and the perpendicular distance to the origin $\rho$ (see Fig.~\ref{fig:samplenetworks}G(i)). The line is then described by the equation $x \sin \left( \theta \right) + y \cos \left( \theta \right) = \rho$. The pair of parameters $(\rho,\theta)$ uniquely defines the line. A Poisson line process samples pairs of parameters from the domain $[-\rho_{\text{max}},\rho_{\text{max}}] \times \left[0,\pi\right)$ at random (see Fig.~\ref{fig:samplenetworks}G(ii)) where $\rho_{\text{max}}$ is a threshold for how far a line we sample is from the origin. Each point in the $(\rho,\theta)$ space corresponds to a line. For example in Figure~\ref{fig:samplenetworks}G(ii) the parameter pair highlighted in green has a $\theta$ coordinate very close to zero, subsequently the corresponding, also highlighted green in Fig.~\ref{fig:samplenetworks}G(iii), is approximately horizontal. Similarly, a parameter pair with $\theta$ coordinate close to $\pi/2$ (see Fig.~\ref{fig:samplenetworks}G(ii), magenta parameter pair) corresponds to a near vertical line. A network can be constructed by sampling Poisson lines and introducing nodes whenever two lines intersect. If the parameter pairs are sampled from an homogenous Poisson point process with intensity parameter $\lambda_{\ell}$ the corresponding network is referred to as an homogenous Poisson line network. The homogenous Poisson line network has no structure in the orientation of the lines. 

\section{Mean first passage times on spatial networks}\label{sec:mfpt}

\begin{figure}
\centering
$\begin{array}{c}
\includegraphics[width=\columnwidth]{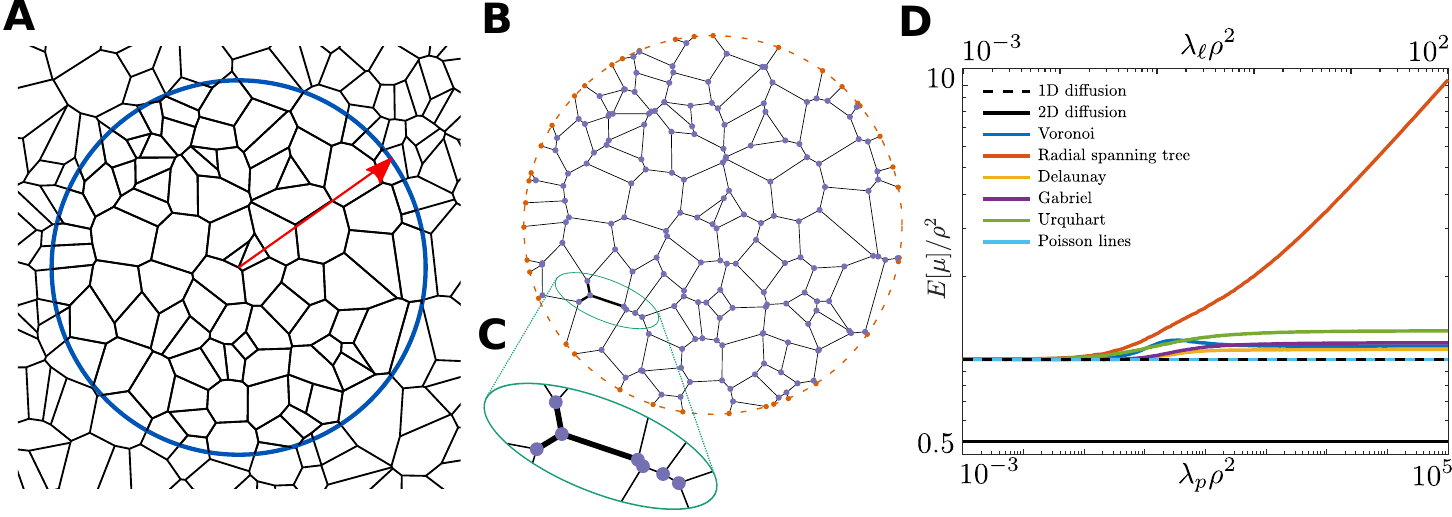}
\end{array}$
\caption{First passage times for a particle on a planar network to reach a given displacement in the plane. (A) An example of a Voronoi network, where a blue circle dictates the positions on the network where a particle will first reach a given displacement, the length of which is given by the circle's radius (red arrow). (B) The intersection of the Voronoi network in panel A and the circle of fixed radius, that provides a finite network for study. New nodes are introduced where the circle and the network intersect and are shown as red dots. (C) Global transport properties on the network in panel B depend upon the local transport properties of arbitrary star graphs (highlighted in thick lines). (D) Mean first passage times normalised by $\rho^2$ for each of the networks in Fig.~\ref{fig:samplenetworks}. The length scales of Poisson point networks are varied on the lower $x$ axis over $200$ different values. For each value of $\lambda_{p} \rho^2$ a total of $10^4$ networks were realised and an average of the mean first passage times for each network was calculated.  The length scales of Poisson line networks are varied on the upper $x$ axis over $751$ different values For each value of $\lambda_{\ell} \rho^2$ a total of $10^2$ networks were realised and an average of the mean first passage times for each network was calculated. The mean first passage times for 1D and 2D off-network diffusion is given by the dashed and solid horizontal lines, respectively.  }
\label{fig:finitenetworks}
\end{figure}

A canonical statistic in the study of random walks is the time at which a walker first reaches a displacement of a given length. This time is known as a first passage time. In this section we will study the affect of network topology on the first passage time on a variety of random walks. First we formalise how such a statistic is calculated. Suppose the initial position of a random walker is at a node on an infinite network $\mathcal{G}_{\infty}$ that spans the plane, and w.l.o.g. let the position of that node be the origin. We are interested in the time taken by the random walker to first reach a displacement $\rho$ from the origin (see Fig.~\ref{fig:finitenetworks} circle). Taking the intersection of the network $\mathcal{G}_{\infty}$ and a circle of radius $\rho$ centered at the origin yields a finite network $\mathcal{G}_{\rho}$. The nodes that lie within the circle are the internal nodes of the finite network $\mathcal{G}_{\rho}$ and if an edge within $\mathcal{G}_{\infty}$ intersects the circle, the point of intersection becomes a new node for the finite network (see Fig.~\ref{fig:finitenetworks}B, red nodes). These new nodes represent all the possible positions in the network $\mathcal{G}_{\infty}$ where a random walker first reaches a displacement $\rho$. Thus, the first passage time to reach a displacement $\rho$ becomes the time taken for a random walker in $\mathcal{G}_{\rho}$ to reach a boundary node. This time can be calculated by studying a random walk on the finite network where the boundary nodes are absorbing.

Let $\mathcal{G}_{\rho} = \left( \mathcal{V}_{\rho} , \mathcal{E}_{\rho} \right)$ denote the finite network and the corresponding set of nodes and edges. The set of boundary nodes are given by $\mathcal{J} \subset \mathcal{V}_{\rho}$. Let $t_{\nu}$ be the time taken for a random walker initially at vertex $\nu \in \mathcal{V}_{\rho}$ to be absorbed by any node in $\mathcal{J}$. By conditioning on the first step of the random walk and invoking the conditional law of expectation we derive the system of linear equations
\begin{equation}\label{eq:LET}
\mathbb{E} \left( t_{\nu} \right) = \sum_{\omega \in \mathcal{V}_{\rho}} p_{\nu \rightarrow \omega} \left[ \mathbb{E} \left( t_{\omega} \right) + \mathbb{E} \left( \tau_{\nu \rightarrow \omega}\right) \right],
\end{equation}	
where $p_{\nu \rightarrow \omega}$ is the probability that the random walker moves from node $\nu$ to $\omega$, and $\tau_{\nu \rightarrow \omega}$ is the time taken for the jump to occur conditioned on the event that the walker moves to node $\omega$. Noting that for $\omega \in \mathcal{J}$ that $\mathbb{E} \left( t_{\omega} \right)=0$ we can partition the sum in Eq.~\eqref{eq:LET} such that
\begin{subequations}\label{eq:LET_partitioned}
\begin{align}
\mathbb{E} \left( t_{\nu} \right) &= \sum_{\omega \in \mathcal{J}} p_{\nu \rightarrow \omega} \mathbb{E} \left(\tau_{\nu \rightarrow \omega}\right) + \sum_{\omega \in \mathcal{V}_{\rho} \setminus \mathcal{J}} p_{\nu \rightarrow \omega} \left[ \mathbb{E} \left(t_{\omega}\right) + \mathbb{E} \left( \tau_{\nu \rightarrow \omega}\right) \right], \\
&= \mathbb{E} \left( \tau_{\nu}\right) + \sum_{\omega \in \mathcal{V}_{\rho} \setminus \mathcal{J}} p_{\nu \rightarrow \omega}  \mathbb{E} \left(t_{\omega}\right), 
\end{align}
\end{subequations}
where $	\mathbb{E} \left( \tau_{\nu}\right) = \sum_{\omega \in \mathcal{V}_{\rho} } p_{\nu \rightarrow \omega} \mathbb{E}\left(\tau_{\nu \rightarrow \omega} \right)$ is the unconditional mean exit time for a random walker to leave node $\nu$. Defining $\vec{T}^{(1)}$ a vector with unknown entries $\mathbb{E}\left( t_{\nu}\right)$, $\vec{\mathcal{T}}$ a vector of known mean exit times $\mathbb{E} \left( \tau_{\nu} \right)$ and $\mathbf{P}$ a matrix with transition probabilities $p_{\nu \rightarrow \omega}$, we can define from Eq.~\eqref{eq:LET_partitioned} a system of linear equations
\begin{equation}\label{eq:LET_Matrix}
\left( \mathbf{I} - \mathbf{P} \right) \vec{T}^{(1)} = \vec{\mathcal{T}},
\end{equation}
where $\mathbf{I}$ is the identity matrix. The linear system of equations defined by Eq.~\eqref{eq:LET_Matrix} are general and can be solved to study any transport process on a network as long as the transition probabilities $p_{\nu \rightarrow \omega}$ between neighbouring nodes and the mean exit times $	\mathbb{E} \left( \tau_{\nu}\right)$ are known. These quantities can be derived for a given transport process by considering first passage properties on the subnetwork of $\mathcal{G}_{\rho}$ consisting of node $\nu$ and their immediate neighbours, this subnetwork is known as a star graph.

\subsection{First passage properties of a star graph: Diffusive transport}

Suppose for a given node within a network, that we term the root node, there are $M$ neighbouring nodes that connect directly to the root node. The star graph is the network consisting of these $M+1$ nodes and the $M$ edges that all connect to the root node (see Fig.~\ref{fig:finitenetworks}C). Let the root node be indexed as $\nu$ and each of the connecting nodes be arbitrarily ordered from $\omega_1$ to $\omega_M$. The edge, that connects the root node to node $\omega_i$, has a length given by the Euclidean distance $\ell_{\nu,\omega_i}$ between the two nodes. We are interested in first passage properties for a particle that occupies the root node. In particular, the probabilities of which neighbouring node is reached first, i.e. the hitting probabilities, as well as the time taken for this absorption to occur, i.e. the first passage time.

We first consider first passage properties on the star graph for a diffusive random walk. Let $p_{\nu \rightarrow \omega_i}$ denote the probability that a particle starting at the root node reaches the $i$-th neighbouring node before any other node. In Appendix A\ref{appendix:stargraph-moments} we detail the derivation of the hitting probabilities $h_i$ by first discretising the edges into $L_i$ lattice sites where $L_i \approx \ell_i / \Delta$, where $\Delta$ is the width of each lattice site. A diffusive random walk with diffusion coefficient $D$ is then formulated on the discretised star graph where the lattice sites corresponding to each neighbouring node are taken to be absorbing nodes. Taking $\Delta \rightarrow 0$ we calculate hitting probabilities $p_{\nu \rightarrow \omega_i}$ in the continuum limit for a diffusive random walk and we find that
\begin{equation}\label{eq:diffusion_hp}
p_{\nu \rightarrow \omega_i} = \dfrac{{\ell_{\nu,\omega_i}}^{-1}}{\sum_{j=1}^{M} {\ell_{\nu,\omega_j}}^{-1}}.
\end{equation}
Furthermore, from the same discretised model (see Appendix) we can calculate the mean time taken for absorption events to occur. Let $\mathbb{E} \left( \tau_{\nu,\omega_i} \right)$ be the mean first passage time for a particle initially at the root node $\nu$ to be absorbed at node $\omega_i$, conditioned on the event that the particle is absorbed by the neighbouring node $\omega_i$. Denoting $h (\vec{\ell})$ and $a(\vec{\ell})$ the harmonic and arithmetic means of the edge lengths $\vec{\ell} = (\ell_{\nu,\omega_1} ,\ldots, \ell_{\nu,\omega_M})$ we find that
\begin{equation}\label{eq:diffusion_cmfpt}
\mathbb{E} \left( \tau_{\nu,\omega_i} \right) = \dfrac{1}{3D} h ( \vec{\ell}) a (\vec{\ell}) + \dfrac{1}{6D} {\ell_{\nu,\omega_i}}^2.
\end{equation}
Noting that the unconditional exit time is given by $\mathbb{E}\left( \tau_{\nu} \right) = \sum_{j=1}^M \mathbb{E}\left( \tau_{\nu,\omega_j} \right) p_{\nu \rightarrow \omega_j}$ we can combine Eqs.~\eqref{eq:diffusion_hp} and \eqref{eq:diffusion_cmfpt} to yield
\begin{equation}\label{eq:diffusion_ucmfpt}
\mathbb{E} \left( \tau_{\nu} \right) = \dfrac{ h ( \vec{\ell}) a (\vec{\ell})}{2D}.
\end{equation}

\subsection{Mean first passage times between spatial networks differ over almost all spatial scales}
The first passage properties of a transport process upon the star graph provides the network transition matrix and vector of mean exit times necessary to solve the linear system in Eq.~\eqref{eq:LET_Matrix}. For a fixed realisation of Poisson points we construct each of the Geometric networks and numerically solve Eq.~\eqref{eq:LET_Matrix} to reveal the mean first passage times. The mean time taken for a random walker to first reach a displacement $\rho$ is unsurprisingly dependent on network topology (see Fig.~\ref{fig:finitenetworks}D). In particular, the Urquhart network is a subgraph of the Gabriel network which is in turn a subgraph of the Delaunay network. Each of the successive subnetworks observes fewer edges than the one before and consequently we see that the mean first passage time rises. Both the radial spanning tree and the minimum spanning tree see significantly greater mean first passage times than the non-tree geometric networks.

In theory as we increase the parameter $\rho$ the circle of radius $\rho$ will include more Poisson points and the networks may reach an equilibrium where the transport process looks like a continuum process, similar to how Cartesian meshes become continuum in the limit of the lattice size tends to zero. The number of points can be increased by either increasing the radius or instead by the intensity of Poisson points by increasing $\lambda_p$. It is in fact a combination of the intensity $\lambda_p$ and the area of the circle that determines how many nodes are in the networks and as such we vary the length scale of our networks by varying the product $\lambda_p \rho^2$. As the length scale increases we expect the first passage times for a diffusive random walker to scale with the squared distance $\rho^2$, thus we normalise the MFPT by $\rho^2$. We find that on small length scales, i.e. $\lambda_p \rho^2 \ll 1$, the rescaled MFPT agrees over all networks and also agrees with the MFPT for a one dimensional random walk. This is because on these very short length scales there are no other nodes in the circle of radius $\rho$ other than the origin node and therefore all networks are in the form of star graphs where every limb equal length. The mean first passage time on these regular star graphs agrees with the MFPT of a one dimensional random walk. On larger length scales the rescaled MFPT no longer agree between networks. For the two types of spanning tree we find that the MFPT does not scale with the squared distance $\rho^2$ and increases without plateau. This suggests that transport on these trees on large length scales does not remain diffusive. This question will be studied significantly later on. The rescaled MFPT for non-tree geometric networks do however reach an equilibrium value. Thus we find that the MFPT for these geometric networks in the large spatial limit scales like $C \rho^2$ however this constant of proportionality $C$ differs between each network.

Let us consider the mean first passage time for a diffusive transport process with diffusion coefficient $D$ in the off-network $d$-dimensional Euclidean space $\mathbb{R}^d$. For a particle initially at the origin, the mean first passage time for the random walker to first reach a displacement of $\rho$ from the origin is $\rho^2 / (2dD)$ (see Appendix A\ref{appendix:off-network-moments}). Note that the first passage time scales quadratically with the distance $\rho$ and both the dimension of the Euclidean space and the diffusion coefficient influence the constant of proportionality. In Figure~\ref{fig:finitenetworks} we have shown the constant of proportionality to be a function of network topology. Naturally we now ask whether the different networks observe a change in diffusion coefficient, a change in effective dimension or both. Fitting the off-network expression for the mean first passage time has one too many degrees of freedom and so we need to consider an additional statistic. The coefficient of variation is defined as the ratio between a random variable's standard deviation and its mean. For the off-network diffusion process the coefficient of variation, $\mathcal{CV}$, is given by
\begin{equation}\label{eq:CoV}
\mathcal{CV} = \sqrt{\dfrac{2}{2+d}}.
\end{equation}
Thus, if we can calculate the coefficient of variation of the first passage time for a given network we can use Eq.~\eqref{eq:CoV} to extract an "effective dimension" for the network. Then use the mean first passage time and the effective dimension to extract the effective diffusion coefficient. However, in order to calculate the coefficient of variation we first need access to higher moments of the networked first passage times.

\section{Higher moments of first passage times}\label{sec:higher-moments}

As for the first moment, we calculate higher moments of first passage times by constructing a linear system of equations to solve numerically. Let $\mathbb{E} \left( t_{\nu}^k \right)$ be the $k$-th moment of the time taken for a particle initially at node $\nu$ to first reaching an absorbing node. Conditioning on the next node that a particle visits yields 
\begin{equation}\label{eq:HigherMomentsConditioning}
\mathbb{E} \left( t_{\nu}^k \right) = \sum_{\omega \in \mathcal{V}_{\rho}} p_{\nu \rightarrow \omega} \mathbb{E} \left[ \left( t_{\omega} + \tau_{\nu \rightarrow \omega}\right)^k\right],
\end{equation}
where as before $p_{\nu \rightarrow \omega}$ is the probability a particle at node $\nu$  reaches node $\omega$ before any other node, and $\tau_{\nu \rightarrow \omega}$ is the time taken for such a transition to occur. Through a binomial expansion and exploiting linearity of expectation we can rewrite Eq.~\eqref{eq:HigherMomentsConditioning} as
\begin{equation}\label{eq:HigherMomentsConditioning2}
\mathbb{E} \left( t_{\nu}^k \right) = \sum_{\omega \in \mathcal{V}_{\rho}} p_{\nu \rightarrow \omega} \sum_{j=0}^k \binom{k}{j} \mathbb{E} \left[ t_{\omega}^j \tau_{\nu \rightarrow \omega}^{k-j}\right],
\end{equation}
and on separating out the $k$-th moment $\mathbb{E} \left[ t_{\omega}^k \right]$ from the internal sum on the right-hand side of Eq.~\eqref{eq:HigherMomentsConditioning2}, we arrive at 
\begin{equation}\label{eq:HigherMomentsConditioning3}
\mathbb{E} \left( t_{\nu}^k \right) = \sum_{\omega \in \mathcal{V}_{\rho}} p_{\nu \rightarrow \omega} \mathbb{E} \left[ t_{\omega}^k \right] + \sum_{\omega \in \mathcal{V}_{\rho}} p_{\nu \rightarrow \omega} \sum_{j=0}^{k-1} \binom{k}{j} \mathbb{E} \left[ t_{\omega}^j \tau_{\nu \rightarrow \omega}^{k-j}\right].
\end{equation}
From Eq.~\eqref{eq:HigherMomentsConditioning3} we can construct the system of linear equations needed to solve for the $k$-th moment of the first passage times. Let $\vec{T}^{(k)}$ be a vector of $k$-th moments $\mathbb{E} \left[ t_{\nu}^k \right]$ for $\nu \in \mathcal{V}_{\rho}\setminus \mathcal{J}$, and let $\vec{\mathcal{T}}^{(k)}$ be a vector with entries $\sum_{j=0}^{k-1} \binom{k}{j} \sum_{\omega \in \mathcal{V}_{\rho}} p_{\nu \rightarrow \omega} \mathbb{E} \left[ t_{\omega}^j \tau_{\nu \rightarrow \omega}^{k-j}\right]$for $\nu \in \mathcal{V}_{\rho}\setminus \mathcal{J}$. With these vectors we can rewrite Eq.~\eqref{eq:HigherMomentsConditioning3} as follows
\begin{equation}\label{eq:KthMomentEquations}
\left( \mathbf{I} - \mathbf{P} \right) \vec{T}^{(k)} = \vec{\mathcal{T}}^{(k)},
\end{equation}
where $\mathbf{I}$ is the identity matrix and $\mathbf{P}$ is the transition matrix. Note that $\vec{\mathcal{T}}^{(1)}$ reduces to the vector of unconditional mean first passage times on the star graph as seen above. Equation \eqref{eq:KthMomentEquations} provides a general set of equations that can be solved numerically for any transport process as long as the transition probabilities $p_{\nu \rightarrow \omega}$ and the expected values $\mathbb{E} \left[ t_{\omega}^j \tau_{\nu \rightarrow \omega}^{k-j}\right]$ are known for $0 \leq j \leq k-1$. In this paper we only consider transport processes where the time to travel between nodes $\nu$ and $\omega$ are independent from the time taken for a particle at node $\omega$ to be absorbed, thus $\mathbb{E} \left[ t_{\omega}^j \tau_{\nu \rightarrow \omega}^{k-j}\right] = \mathbb{E} \left[ t_{\omega}^j \right] \mathbb{E} \left[\tau_{\nu \rightarrow \omega}^{k-j}\right]$. Therefore, solving for the $k$-th moments of the first passage time requires knowledge of the all the proceeding moments $\mathbb{E} \left[ t_{\omega}^j \right]$ for $1 \leq j \leq k-1$ as well as the moments of the first passage time on a star graph, $\mathbb{E} \left[\tau_{\nu \rightarrow \omega}^{j}\right]$ for $1 \leq j \leq k$.
\subsection{Second moment first passage time on a star graph: Diffusive transport}
In order to calculate the second moment of a networks first passage time for a diffusive transport process we need to calculate the second moment of the first passage times on a star graph. From expanding the right hand side of Eq.~\eqref{eq:KthMomentEquations} for $k=2$ we see that we only need the unconditional second moment for the first passage times $\mathbb{E} \left[\tau_{\nu}^{2}\right]$. Appendix A\ref{appendix:stargraph-moments} presents a similar derivation to the first moment, $\mathbb{E} \left[\tau_{\nu}\right]$, and we find that
\begin{equation}
\mathbb{E} \left[\tau_{\nu}^{2}\right] = \dfrac{1}{3D^2} \left( h(\vec{\ell}) a(\vec{\ell}) \right)^2 + \dfrac{1}{12D^2} \left( h(\vec{\ell}) a(\vec{\ell}^{\ 3}) \right),
\end{equation}
where $a(\vec{\ell})$ and $h(\vec{\ell})$ are the arithmetic and harmonic means of the lengths of the limbs in the star graph, and $a(\vec{\ell}^{\ 3})$ is the arithmetic mean of the cubed limb lengths.
\subsection{Coefficient of variation and effective dimensionality}
\begin{figure}
\centering
$\begin{array}{c}
\includegraphics[scale=0.8]{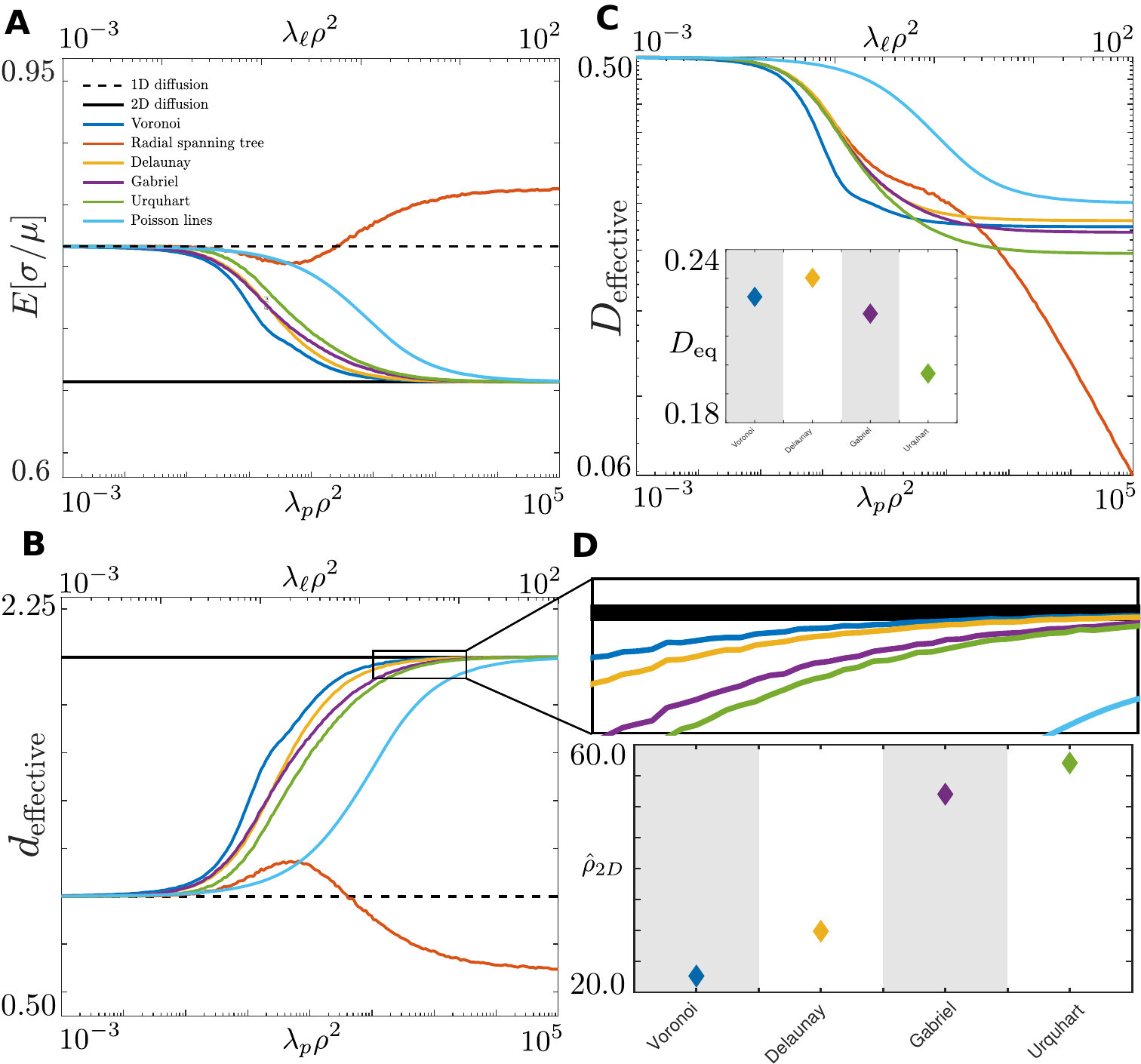}
\end{array}$
\caption{The coefficient of variation reveals the length scales upon which networked transport does or does not transition to two-dimensional off-network diffusion. (A) Coefficient of variation as a function of length scale for all networks seen in Fig.~\ref{fig:samplenetworks}. The coefficient of variation for first passage times for 1D and 2D off-network diffusion is given by the dashed and solid horizontal lines, respectively. (B) Effective dimension of networked transport as a function of length scale for all networks seen in Fig.~\ref{fig:samplenetworks}. (C) Effective diffusion coefficient of of networked transport as a function of length scale for all networks seen in Fig.~\ref{fig:samplenetworks}. The inset shows the effective diffusion coefficient at the largest spatial scale for each of the networks that see convergence. (D) For each of the networks that see a convergence to two-dimensional off-network diffusion the length scale on which this transition occurs is given the length scale at which the effective dimension hits $1.99$. For (A-D) the length scales of Poisson point networks are varied on the lower $x$ axis over $200$ different values. For each value of $\lambda_{p} \rho^2$ a total of $10^4$ networks were realised and an average of the mean first passage times for each network was calculated.  The length scales of Poisson line networks are varied on the upper $x$ axis over $751$ different values For each value of $\lambda_{\ell} \rho^2$ a total of $10^2$ networks were realised and an average of the mean first passage times for each network was calculated.}
\label{fig:CoV}
\end{figure}
On short length scales (i.e. $\lambda \rho^2 \ll 1$) we see that the coefficient of variation unsurpisingly agrees with one-dimensional off-network diffusion (see Fig.~\ref{fig:CoV}A). As this length scale is increased the coefficient of variation falls for all networks until plateauing at the coefficient of variation for two-dimensional off network diffusion (see Fig.~\ref{fig:CoV}A), except for the radial spanning tree which is discussed later. This transition in effective dimensionality is confirmed by using the coefficient of variation to extract the effective dimension from Eq.~\eqref{eq:CoV} (see Fig.~\ref{fig:CoV}B). This confirms our intuition that as the density of nodes increases, or equivalently the length scale is increased, networked diffusion will usually transition to two-dimensional Euclidean diffusion. The differences in network structure are then encapsulated by the effective diffusion coefficient $D_{\text{effective}}$, that can be extracted from fitting the MFPT to $\rho^2/(2d_{\text{effective}}D_{\text{effective}})$ (see Fig.~\ref{fig:CoV}C). The diffusion coefficients on large length scales are a function of the connectivity of the different networks. The Delaunay, Gabriel and Urquhart networks which are all subnetworks of the previous one have decreasing effective diffusion coefficients reflecting the reduction in connectivity. It is clear from Figure~\ref{fig:CoV}B that there are length scales on which first passage times are still transitioning between effectively one and two dimensional diffusion. To quantify the length scales on which these networks appear two-dimensional we introduce $\hat{\rho}_{2\text{D}}$ as the first value of $\rho$ such that the effective dimension equals $1.99$ (see Fig.~\ref{fig:CoV}D). Network structure has a large role to play in the size of this length scale. Indeed, for an Urquhart network the length scale is almost three times larger than for a Voronoi network. However, not all networks that we considered appear to converge to standard two-dimensional diffusion behaviour.

The coefficient of variation for the RST does has non-monotonic behaviour with the length scale parameter $\lambda \rho^2$ and does converge to the effective two-dimensional result. As such we cannot use results from standard Euclidean diffusion to extract effective parameters. However, as we saw that the MFPT for the RST did not scale with $\rho^2$ this was to be expected. We hypothesise that the unusual behaviour of first passage times on RST is due to the tree structure of this network. To test our hypothesis we design a new network structure that perturbs the RST away from being a tree.

\subsection{Perturbed radial spanning tree}

\begin{figure}
\centering
$\begin{array}{c}
\includegraphics[width=\columnwidth]{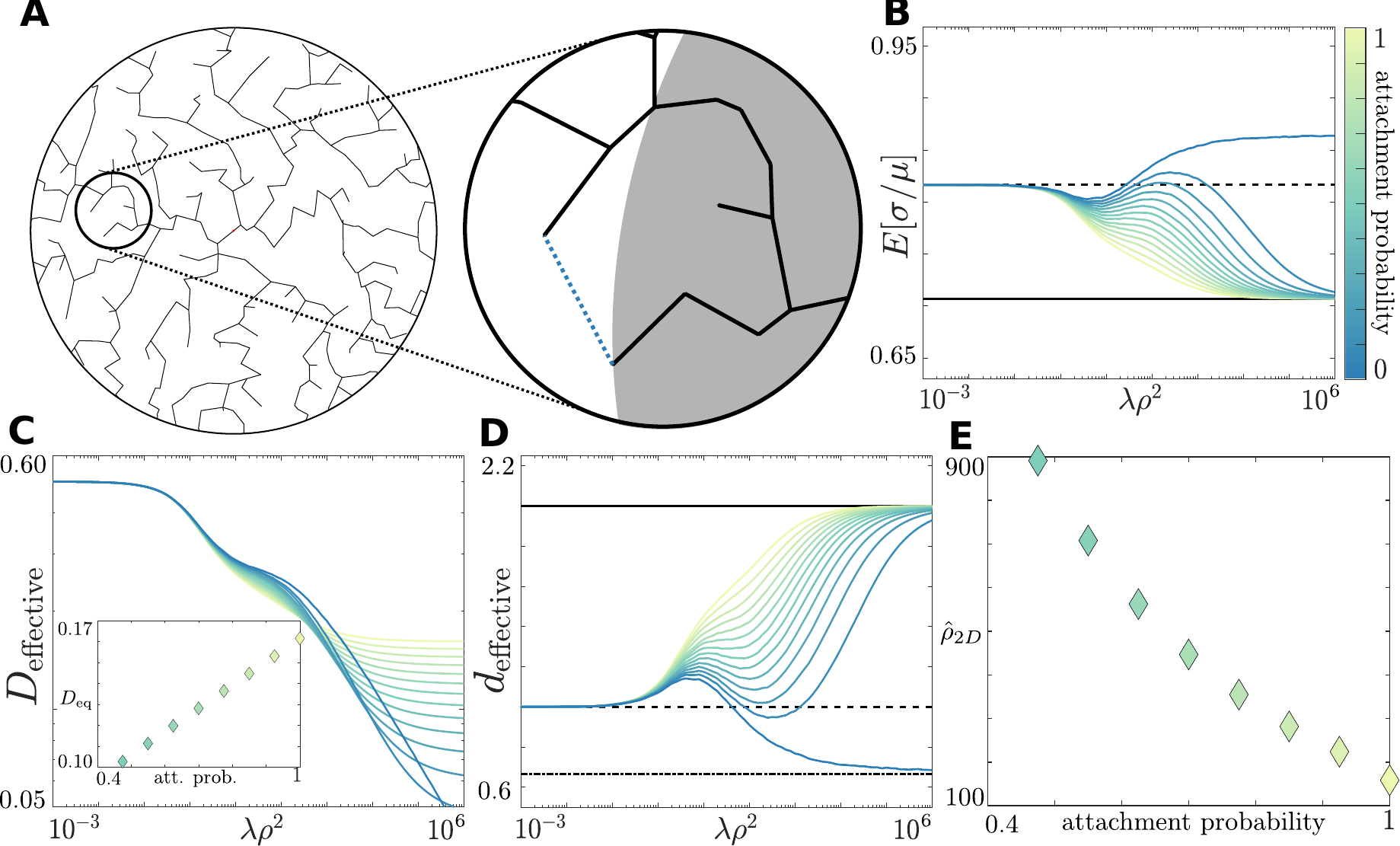}
\end{array}$
\caption{Perturbing the radial spanning tree shows 'tree-like' networks will converge to two-dimensional planar diffusion but over very large length scales. (A) An example of a radial spanning tree, the inset shows an edge that connects two stubs (blue dashed line) creating a perturbed radial spanning tree. (B) Coefficient of variation as a function of length scale for transport on perturbed radiall spanning trees for a range of attachment probabilites shown by the colourbar. The mean first passage times for 1D and 2D off-network diffusion is given by the dashed and solid horizontal lines, respectively. (C) Effective diffusion coefficient as a function of length scale for transport on perturbed radiall spanning trees for a range of attachment probabilites shown by the colourbar. The inset shows the effective diffusion cofficient for the largest spatial scale as a function of the attachment probability. (D) Effective dimension as a function of length scale for transport on perturbed radiall spanning trees for a range of attachment probabilites shown by the colourbar. (E) The length scale on which transport upon perturbed radial spanning trees with non-zero attachment probabilities converge to two-dimensional off-network diffusion, as defined by the length scale they reach an effective dimension of $1.99$. The plots in (B-E) for $14$ different attachment probabilities and $91$ different length scales $\lambda \rho^2$, and for each pair of length scale and attachment probability $10^4$ perturbed radial spanning trees were sampled. 	 }
\label{fig:PRST}
\end{figure}

A perturbed radial spanning tree (PRST) is built by first constructing a regular RST and subsequently adding edges to the ends of some of the leaves (nodes of degree one). Thus creating a non-tree network that still appears visually very similar to a RST. The way we select the additional edges to include is as follows. For every leaf in the RST (see Fig.~\ref{fig:PRST}A, starred node) we add an edge with a probability $p_{\text{attachment}}$ which we term the attachment probability. The leaf is then connected to the nearest node that is farther away from the root node (see Fig.~\ref{fig:PRST}A, excluded region in the inset). Ensuring that the node is farther away from the root node minimises the possibility of the network becoming non-planar. If the new edge does intersect with another edge it is rejectd. These new edges are sampled for each leaf starting from the leaf closest to the root and working outwards. We introduce the attachment probability such that we can explore networks that are very similar to the RST but are not trees, i.e. $0<p_{\text{attachment}}\ll1$.

For $p_{\text{attachment}}=1$, where all leaves are connected and there are no more nodes of degree one, we find that the network behaves like the other non-tree networks we have considered (Voronoi, Delaunay, etc.). The relationship between the length scale and the coefficient of variation is monotonic and the effective dimension tends towards two. However as the attachment probability is reduced the relationship between length scale and coefficient of variation of the first passage time becomes non-monotonic as for the RST. However, on large enough length scales the behaviour does tend towards being two-dimensional. Therefore we have additional numerical evidence that it is the tree property of the RST that is causing the deviation away from standard diffusion. Although the rare occurrence of new edges (i.e. $0<p_{\text{attachment}}\ll1$)  does completely change the transport behaviour to the RST the length scales upon which the transition to two dimensional behaviour are very large. For $p_{\text{attachment}} \approx 0.5$ we find that $\hat{\rho}_{2\text{D}}$ is about $15$ times larger (see Fig.~4E) than for the Urquhart network (the least connected network we considered previously). This is particularly interesting as many naturally occurring biological networks are "tree-like" but not complete trees.

\section{Radially dependent diffusion coefficient}\label{sec:radial-diffusion}

\begin{figure}
\centering
$\begin{array}{c}
\includegraphics[scale=0.8]{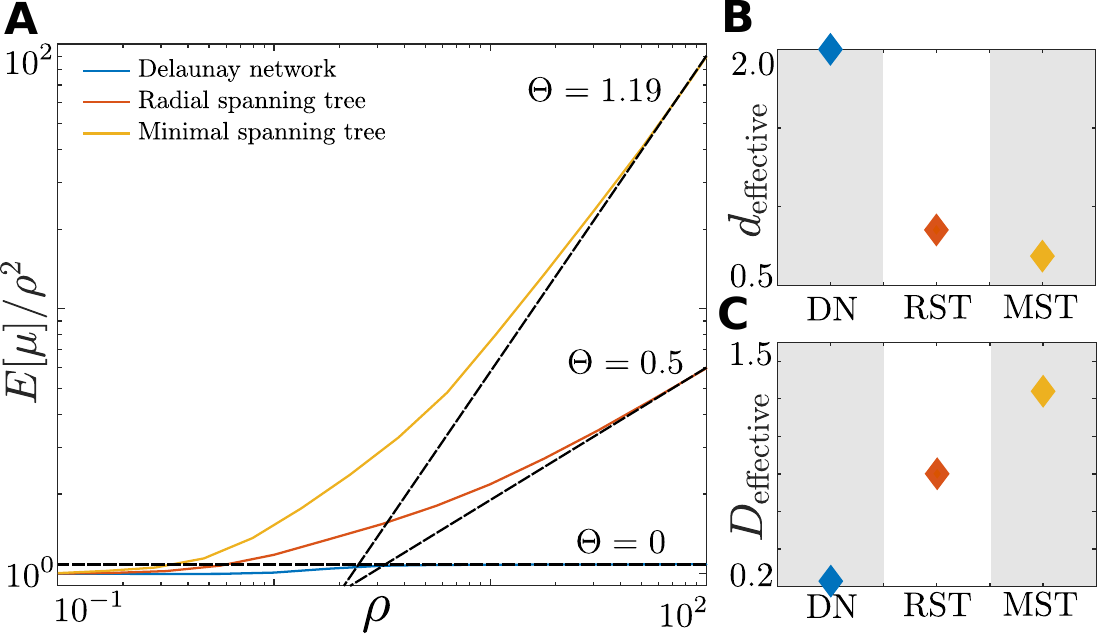}
\end{array}$
\caption{Radially dependent diffusion can describe the first two moments of first passage times for diffusive transport on trees. (A) The mean first passage time normalised by $\rho^2$ as a function of $\rho$ for Delaunay networks, radial spanning trees and minimum spanning trees. The exponent of the diffusion coefficient $\Theta$ is extracted by the gradient of the logarithm of the mean first passage time curves (black dashed lines). The radius $\rho$ is varied over $17$ different values and for each value $10^4$ networks were sample and an average of their mean first passage times is taken. (B) The effective dimension for transport averaged over $10^4$ realisations of each of the networks where $\lambda=1$ and $\rho=10^2$. (C) The effective diffusion coefficient for transport averaged over $10^4$ realisations of each of the networks where $\lambda=1$ and $\rho=10^2$. }
\label{fig:radial-diffusion}
\end{figure}

When comparing networked transport to standard planar diffusion, the RST produced diffusion coefficients that diverged for larger and larger length scales (Fig.~\ref{fig:CoV}C). This suggests that the effective diffusion coefficient for a particle diffusing on the RST is a function of the displacement distance $\rho$. For a RST, a node is selected as the root node, and the farther from the root node the more branches appear in the tree. Intuitively, more branches will mean more dead-ends and the effective diffusion coefficient of the particle would be reduced. To investigate transport along RSTs further we consider a planar diffusion process where the diffusion coefficient is given by $D\left( \rho \right) = D \rho^{-\Theta}$. This generalised diffusion process has been used in previous studies to investigate diffusion within fractal environments \cite{OShaughnessy_PRL85}. In Appendix~A\ref{appendix:off-network-moments} we consider the first passage time problem for a particle undergoing radially dependent off-network diffusion. Our calculations reveal that the first moment of the first passage time is given by $\rho^{2+\Theta} / \left( \left( 2 + \Theta \right) d D \right)$. Thus, for radially dependent diffusion coefficients we see a rise in the exponent. This agrees with RST where we saw in Figure \ref{fig:finitenetworks}C that the mean FPT for the RST did not scale quadratically. Using the networked mean FPT for the RST we can numerically extract the value of $\Theta$ and we find that $\Theta = 0.5$ (see Fig.~\ref{fig:radial-diffusion}A). For a minimum spanning tree (this network connects all the Poisson points whilst minimising the total sum of all the edges) we find that the exponent is higher at about $\Theta = 1.19$ (see Fig.~\ref{fig:radial-diffusion}A). We wish to calculate the effective dimension and diffusion coefficient, so just as for standard diffusion we derive an expression (Appendix~A\ref{appendix:off-network-moments}) for the coefficient of variation
\begin{equation}
\mathcal{CV} = \sqrt{\dfrac{2+\Theta}{2+\Theta+d}}.
\end{equation}
The coefficient of variation now depends on both the exponent $\Theta$ and the effective dimension $d$. As we extracted the exponent from studying the first moment we can use the coefficient of variation to extract the effective dimension (see Fig.~\ref{fig:radial-diffusion}B). For the RST we find an effective dimension of approximately $0.85$ and for the MST we find an effective dimension of approximately $0.69$. Note that these dimensions are both non-integer and less than one. Thus contrary to the Delaunay network (and all the other non-tree networks) it appears the effective dimension of the transport on the spanning trees are reduced on larger spatial scales and do not align with an integer dimension. Interesting the non-integer dimensionality does not lie between one and two dimensions but below one (@Casper is this a weird sentence, i dont have much more to say). Finally, from the equation for the mean FPT, $\rho^{2+\Theta} / \left( \left( 2 + \Theta \right) d D \right)$, we can extract the effective diffusion coefficient by fitting the last unknown parameter $D$ (see Fig.~\ref{fig:radial-diffusion}C). 

Thus far, we have shown that it is possible to relate an off-network diffusion process with either constant or radially dependent diffusion coefficient that agrees with the mean and coefficient of variation of first passage times for a diffusive particle on a network. However, our analysis has only considered the first two moments of the first-passage times. Two moments are sufficient to identify the mapping to effective planar diffusion, as planar diffusion is uniquely defined by the first two moments of the FPT. However, we need to check the higher moments of the networked first-passage times to make sure they still agree with the planar diffusive processes.

\section{Do networked FPTs agree with planar diffusion FPTs for all moments}\label{sec:skewness-kurtosis-distributions}


Considering higher moments of the first passage time of a particle on a network will require calculating higher first passage times on a star graph. For diffusive random walkers this is a particularly cumbersome task. As such, here we consider an alternative transport process, namely ballistic transport. For this process all   moments of first passage times on the star graph are immediately available.


\subsection{First passage properties of a star graph: Ballistic transport}

Here we consider an example of an active transport process where a random walker at the root node of a star graph selects a limb to travel along at a constant speed $V$. The limb that the walker chooses is selected uniformly at random, which yields uniform hitting probabilities $p_{\nu \rightarrow \omega_i} = 1/M$. Once a random walker selects the limb connecting nodes $\nu$ and $\omega_i$ the first passage time becomes deterministic and is given by $\tau_{\nu,\omega_i} = \ell_{\nu,\omega_i}/V$. Indeed all moments of the first passage time on a star graph is given by $\mathbb{E} \left( \tau_{\nu \rightarrow \omega_i}^k \right) = \left( \ell_{\nu,\omega_i} / V \right)^k$. Thus we have all the information needed to calculate any moment of the first passage time to reach a given displacement $\rho$. 

\subsection{Skewness and Kurtosis}


Given all moments for first passage times on a star graph we can calculate from Equations~\eqref{eq:KthMomentEquations} arbitrarily many moments of networked first passage for a ballistic transport process. The mean and the coefficient of variation of networked first passage times can be mapped to effective planar diffusion first passage times on large spatial scales\footnote{For brevity, these results are not shown in the paper.}, just as for the diffusive random walker. This is expected as ballistic transport on a network has similar properties to a velocity jump model (a transport process in the plane where a particle moves in a direction at a constant speed and after an exponentially distributed amount of time reorients and selects a new direction) which is proven to be diffusive over large spatial scales. However, in order to gain further numerical evidence that networked first passage times are the same as planar diffusive first passage times we consider two more summary statistics. Let $T$ be the first passage time then the standardized central moments of $T$ are given by $S_k = \mathbb{E}\left[ \left(X -\mathbb{E} (X) \right)^k \right]  \mathbb{E}\left[ \left(X -\mathbb{E} (X) \right)^2 \right]^{-k/2}$. Note that $S_2$ is equal to the coefficient of variation. The two new summary statistics we consider are $S_3$ adn $S_4$ which are known as the skewness and kurtosis of the first passage time.

\begin{figure}
\centering
$\begin{array}{c}
\includegraphics[width=\columnwidth]{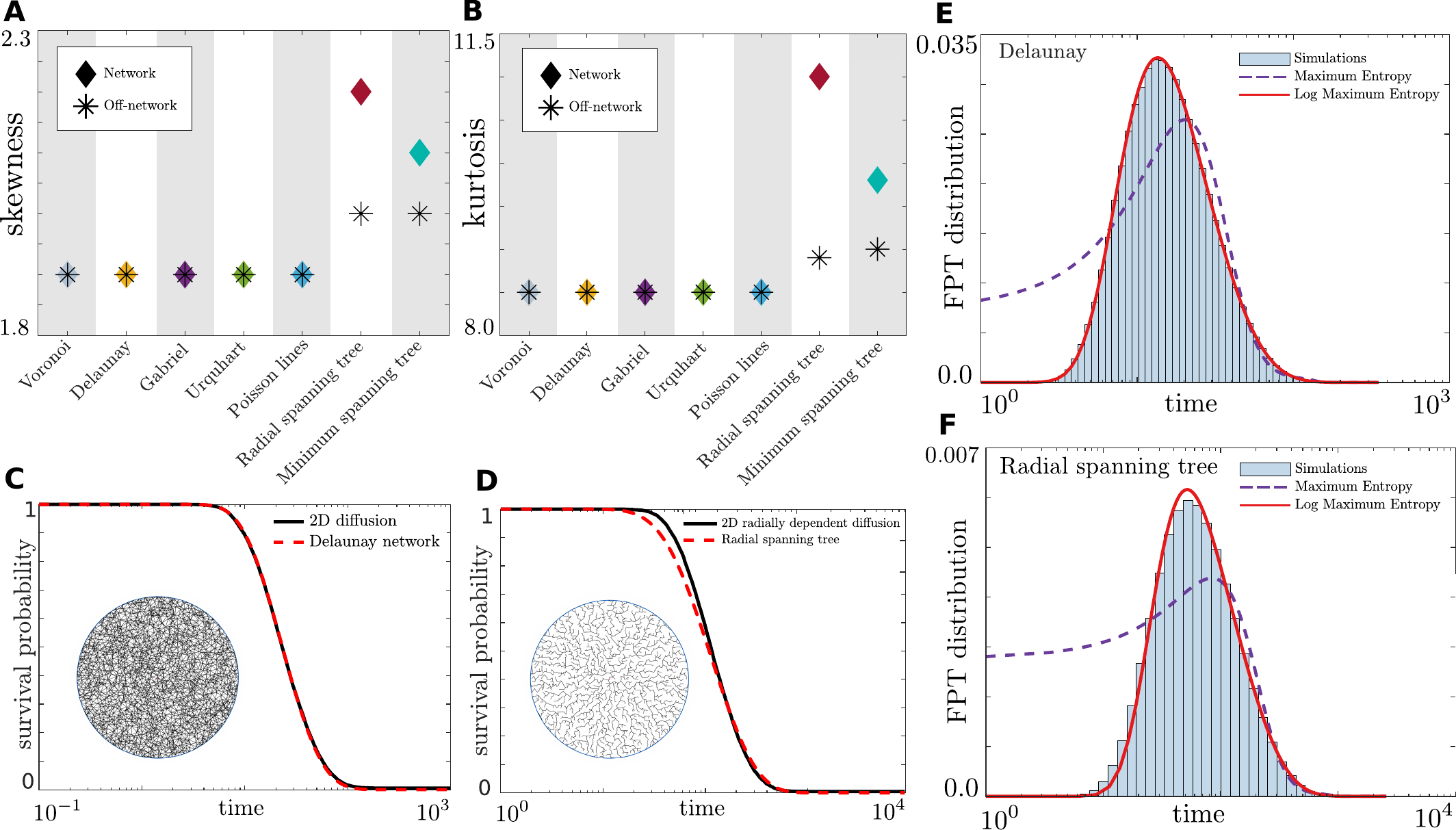}
\end{array}$
\caption{Higher moments and entire distributions of networked first passage times. (A) The skewness for first passage times averaged over $10^4$ realisations of each of the networks where $\lambda=1$ and $\rho=10^2$ The skewness for first passage times on off-network diffusion processes, both classical and radial, are shown in black asterisks. (B) The kurtosis for first passage times averaged over $10^4$ realisations of each of the networks where $\lambda=1$ and $\rho=10^2$ The kurtosis for first passage times on off-network diffusion processes, both classical and radial, are shown in black asterisks. (C) The survival probability for transport on a single Delaunay network (inset) with $\lambda=1$ and $\rho=10^2$ (red dashed curve), compared with the first passage time of a Brownian walker with diffusion coefficient extracted from the Delaunay network analysis (black solid curve). Both these survival probabilities were estimated emperically from $10^6$ stochastic realisations of the transport processes. (D) The survival probability for transport on a single radial spanning tree (inset) with $\lambda=1$ and $\rho=10^2$ (red dashed curve), compared with the first passage time of a radially dependent Brownian walker with diffusion coefficient and exponent extracted from the radial tree analysis (black solid curve). Both these survival probabilities were estimated emperically from $10^6$ stochastic realisations of the transport processes. (E) First passage time distribution for transport for the Delaunay network seen in panel C (histogram). Compared with both the maximum entropy distribution (purple dashed curve) and log maximum entropy distribution (red solid curve), reconstucted using the first four moments. (F) First passage time distribution for transport for the radial spanning tree seen in panel D (histogram). Compared with both the maximum entropy distribution (purple dashed curve) and log maximum entropy distribution (red solid curve), reconstucted using the first four moments. The histograms in (E-F) were estimated from $10^6$ stochastic realisations of networked ballistic transport. }
\label{fig:6}
\end{figure}

For the Voronoi, Delaunay, Gabriel, Urquhart and Poisson line networks we find that the skewness and kurtosis of the first passage times on networks on large spatial scales agree with the skewness and kurtosis of the first passage times of a particle diffusing on the plane (see Fig.~\ref{fig:6}A,B and Appendix\ref{appendix:off-network-moments}(iv)). Thus there is numerical evidence suporting the claim that  first passage times for non-tree networked transport processes will converge to first passage times for planar diffusive processes in the large spatial limit. To provide even further evidence we consider the survival probablity (defined as the probability a particle intially at the origin has not been absorbed by time $t$) for a particle undergoing ballistic transport on a single realisation of a large Delaunay network (see Fig.~\ref{fig:6}C, inset). Comparing the empirically estimated survival probability on the Delaunay network with the survival probability with planar diffusion reveals that they are infact equal (see Fig.~\ref{fig:6}C). As the survival probability is equivalent to the first passage time distribution (calculated by the taking the negative derivative with respect to time), we conclude that we believe first passage times for non-tree 2D planar networked transport processes will converge to first passage times for planar diffusive processes in the large spatial limit.

However, for the RST and MST the skewness and kurtosis for networked first passage times do not agree with the first passage times of planar radially dependent diffusion (see Fig.~\ref{fig:6}A,B). Thus, whilst the first two moments of the first passage time for transport on trees can be mapped to an effective radially dependent planar diffusion process, the disagreement at higher moments reveals that there can be no convergence as is seen for the other networks. However the  emperical estimates of the survival probability for a single realisation of a RST (see Fig.~\ref{fig:6}D, inset) and a radially dependent planar diffusion process, that show there is only a slight disagreement between the two (see Fig.~\ref{fig:6}D). 

\subsection{Reconstructing first passage time distributions}

We have already stated that our methodology can produce arbitrarily many moments of the first passage time, as long as the moments of the star graph first passage problem are known. In the case of ballistic motion this is trivial. Given that we have access to all moments numerically we can attempt to use these moments to reconstruct the entire distribution of the first passage time for any network without having to simulate millions of stochastic simulations. How one should reconstruct a distribution using a finite number of moments remains a question of active research. A common approach is to use a maximum entropy distribution \cite{Wilson_AIP16}. In essence this approach selects from all the distributions with the prescribed moments the distribution that is "most likely" . When reconstructing a distribution with $M$ moments, the probability distribution function, $p_{\text{ME}}(t)$ of a maximum entropy distribution takes the following form,
\begin{equation}
p_{\text{ME}}(t) = \exp \left( - \sum_{m=0}^M \lambda_m t^m\right),
\end{equation}
where $\lambda_m$ are unknown constants that we select in order to achieve our prescribed moments. Put another way, given a finite set of moments, $\left\lbrace \mu_0,\mu_1,\ldots,\mu_M \right\rbrace$ we need to calculate the constants $\lambda_m$ such that
\begin{equation}
\mu_m = \int_{0}^{\infty} t^m p_{\text{ME}}(t) \mathrm{d}t.
\end{equation}
The numerical method used to solve these integral equations, along with available MATLAB code, can be found in \cite{Djafari_MATLAB}. 

From Equations~\eqref{eq:KthMomentEquations} we can solve for the first $M$ moments of the first passage time for a given network and then calculate the maximum entropy distribution. However, this offers a poor approximation. For ballistic transport we simulated one million realisations of the first passage time for a single Delaunay network (see Fig.~\ref{fig:6}E). The maximum entropy distribution matches the empirical distribution at the tail, however for short times the two distributions show poor agreement. This is due to the steep initial incline of the distribution that is difficult to match to a maximum entropy distribution. However, the logarithm of the first passage time shows empirically a much smoother unimodal distribution. Such distributions are ideal for approximation by a maximum entropy distribution. Therefore we make the assumption that $\log(T)$, where $T$ is the first passage time, approximately follows a maximum entropy distribution. We note therefore that the pdf of $T$ is given by $p_T(t) = t^{-1} p_{\text{ME}}(\log(t))$. We therefore make the assumption that the first passage time has the following distribution.
\begin{equation}\label{eq:logME}
p_T(t) = \dfrac{1}{t} \exp \left( - \sum_{m=0}^M \lambda_m \log(t)^m\right).
\end{equation}
Now we solve for the constants $\lambda_m$ using the same approach as for the maximum entropy distribution (see \cite{Djafari_MATLAB}). Using Eq.~\ref{eq:logME} gives an excellent analytical approximation to the first passage time distribution for both the Delaunay network and the RST (see Fig.~\ref{fig:6}E,F). Thus without the need to run a single stochastic simulation we can provide analytical approximations for first passage time distributions on arbitrary networks.

\section{Discussion}\label{sec:discussion}


In this work we have introduced a hierarcichal system of equations that can be solved to obtain all moments of FPTs for diffusing particles on spatially embedded networks to first reach a fixed displacment within the plane. These equations are solveable for all Markovian transport processes as long as the first passage properties of a particle on the arbitrary star graph are well understood. We provided the first passage properties of diffusion on a star graph to allow for the study of diffusion in spatially embedded networks. For a variety of stochastic spatial networks we matched the mean and coefficient of variation of FPTs to the FPTs of off-network planar diffusion. Connecting networked transport processes with planar diffusion revealed how different networks result in different effective diffusion coefficients. We also quantified the length scales upon which diffusive transport on planar networks will transition from one-dimensional to two-dimensional diffusion. However, not all networks, in particular two types of tree, could have their transport matched to classical diffusion. Instead, we showed that the mean and coefficient of variation for FPTs of diffusing particles on RSTs and MSTs could be mapped to radially dependent off-network diffusion. We used our framework to study FPTs for diffusing particles on a class of networks that perturb away from being tree, which we termed 'tree-like'. Our results suggest that tree-like networks on large enough length scales will become effectively classically diffusive, how the length scales required for convergence are incredibly large. Finally we considered a new transport process, ballistic transport, where we had access to all moments of FPTs on the star graph. By exploring higher moments of networked FPTs we showed how to use an adaptation of maximum entropy arguments to calculate analytical approximations of networked FPTs.


There ample opportunities for interesting extensions of this work. In particular, extending the study of first passage properties on arbitrary star graphs to other transport processes, such as velocity jump processes \cite{Hillen_VJP} and biased diffusion \cite{Gefen_JPAMG}, would immediately grant access to FPTs of these transport processes on entire spatial networks. A noteable example for extension would be to incorporate crowding affects with other moelcules withing the network. Recent work showed that the introducing crowding affects to diffusive particles in networked topologies drastically changes the paths taken by particles navigating the network \cite{WilsonCommPhys21}. Recall that for higher moments of networked FPTs we assumed that the time taken for a particle to move between two nodes $\nu$ and $\omega$ is independent from the time taken to subsequently leave $\omega$, i.e.$\mathbb{E} \left[ t_{\omega}^j \tau_{\nu \rightarrow \omega}^{k-j}\right] = \mathbb{E} \left[ t_{\omega}^j \right] \mathbb{E}\left[ \tau_{\nu \rightarrow \omega}^{k-j}\right]$. Future work will explore the potential of extending this framework to non-Markovian transport processes \cite{MetzlerKlafter} where instead the covariance between these two times would need to be calculated explicitly. Overall, this work contributes a highly efficient numerical method to study FPTs of various transport processes over a huge range of different  networks. Circumventing the need to simulate transport processes explicitly allows for studies of FPTs that look at hundreds of thousands of networks, and has the potential to reveal new insights into the interplay between network topology and transport behaviour.

\section*{Acknowledgements}

The authors would like to thank both Prof. Baker Dr. Francis Woodhouse for enlgihtening and insightful discussions about this work. This work was supported by the EPSRC Systems Biology DTC Grant No. EP/G03706X/1 (D.B.W.) and NSF-DMS, 1902854 (D.B.W.).

\section*{Author Contributions}
All authors contributed at all stages of this work.

\bibliography{scibib}

\newpage
\appendix
\section{First-passage times for diffusion on an arbitrary star graph}\label{appendix:stargraph-moments}
\subsubsection{Discrete random walk model}

\begin{figure}
\centering
$\begin{array}{c}
\includegraphics[scale=0.4]{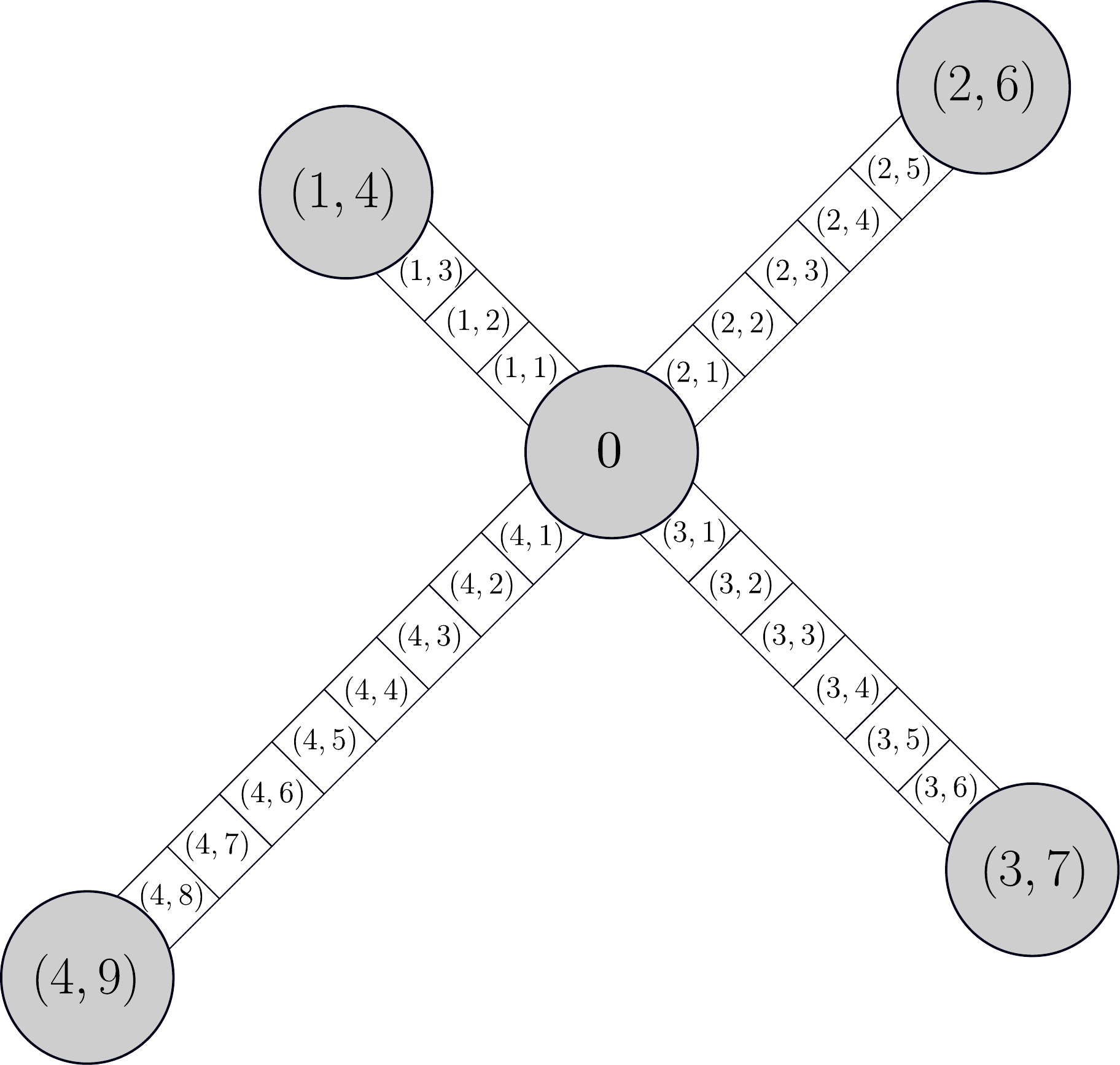}
\end{array}$
\caption{A discretisation of a four-limbed star graph with lengths $L_1=4$, $L_2=6$, $L_3=7$ and $L_4=9$.}
\label{fig:StarGraph}
\end{figure}

Consider a root node with index $0$, this node is connected with $M$ limbs of a star graph with integer lengths $L_1,\dots,L_M$. The vecotr of limb lengths is denoted $\vec{L}=(L_1,\dots,L_M)$. We index the nodes at the end of each limb $(i,L_i)$, see Fig.~\ref{fig:StarGraph} for an example diagram. A particle is initally located at the root node $0$ at time $0$, then aftera time step of length $1$ a particle chooses a new neighbouring site uniformly at random. Let $n$ be the number of steps until the particle is absorbed at the end of any of the limbs of the star graph.

\paragraph{Absorption probability}

First we consider the conditional absorption probability, w.l.o.g.\ assuming that we are interested at absorption at end of limb 1. Following Wilson et al.\ define the conditional hitting probabilities
\begin{subequations}
\begin{align}
    h_{i,j} &= \mathbb{P}[\text{ absorbed at } (1,L_1)\ | \text{ starting from }(i,j)\ ], \\
    h_0 &= \mathbb{P}[\text{ absorbed at } (1,L_1)\ | \text{ starting from the origin }],
\end{align}
\end{subequations}
where we identify $h_{i,0}\equiv h_0$ for all $1\leq i \leq M$. By conditioning on the next move of the walker on the star graph we arrive at the following system of algebraic equations describing the conditional hitting probabilities
\begin{subequations}
\begin{align}
h_{1,L_1} & = 1, & \\
h_{i,L_i} & = 0, &\text{for }2\leq i\leq M, \\
h_0 &= \frac{1}{M} \sum_{i=1}^M h_{i,1}, & \\
h_{i,j} &= \frac{1}{2} h_{i,j-1} + \frac{1}{2}h_{i,j+1}, & \text{for }1\leq i\leq M\text{ and }1\leq j \leq L_i-1.
\end{align}
\end{subequations}	

Solving these recurrence relations yields the hitting probabilities for each limb $i$ and $1\leq j \leq L_i$
\begin{equation}\label{eq:absorption_prob}
    h_{i,j} = \begin{cases} h_0(1-j/L_i), &\qquad \text{if } i\neq 1\\
        h_0(1-j/L_1) + j/L_1, &\qquad \text{if } i=1
    \end{cases},
\end{equation}
where $h_0$ is the hitting probability starting from the origin. This can be found to be
\begin{equation}
    h_0 = L_1^{-1} \left(\sum_{m=1}^M L_m^{-1}\right)^{-1} = \frac{h(\vec{L})}{M L_1},
\end{equation}
where $h(\vec{L})=M/(\sum_m L_m^{-1})$ is the harmonic mean of the limb lengths $L_m$. 

\paragraph{Mean absorption time} First we consider the problem of finding the mean number of steps till absorption at any of the limb ends, defined as
\begin{subequations}
\begin{align}
k_{i,j} &= \mathbb{E}[\ n \ |\ \text{ starting from } (i,j)\ ],\\
k_0 &= \mathbb{E}[ \ n \ |\ \text{ starting from the origin}\ ],
\end{align} 
\end{subequations}
where again we identify $k_{i,0}\equiv k_0$ for all $1\leq i\leq M$. Similar to the previous section we use conditioning on the next move of the walker to derive the set of algebraic equations describing the $k_{i,j}$
\begin{subequations}
\begin{align}
k_{i,L_i} & = 0, &\text{for }1\leq i\leq M, \\
k_0       &= 1 + \sum_{i=1}^M \frac{1}{M}k_{i,1}, \\
k_{i,j}   &= 1 + \frac{1}{2} k_{i,j-1} + \frac{1}{2}k_{i,j+1}, & \text{for }1\leq i\leq M\text{ and }1\leq j \leq L_i-1.
\end{align}
\end{subequations}	
This system is readily solved by 
\begin{subequations}\label{eq:mean_star_graph}
\begin{align}
k_0 &= h(\vec{L}) a(\vec{L}),\\
k_{i,j} &= \left(1-\frac{j}{L_i}\right)\left(k_0 + j L_i\right), \quad \text{for }1\leq j \leq L_i,
\end{align} 
\end{subequations}
where $a(\vec{L})=M^{-1}\sum_m L_m$ is the arithmetic mean of the limb lengths $L_m$.

Finally we consider the (harder) problem of finding the conditional mean absorption time of the walker on the star graph. To this end we introduce the conditional random walker on the star graph, i.e.\ we look at the random walker conditioned on being absorbed at the end of a specific limb, which we take again w.l.o.g.\ to be limb 1. Let $v$ and $w$ be two vertices in the star graph with transition probability for the regular random walker $p_{vw}=\mathbb{P}[v \to w]$.
The transition probabilities for the biased random walk (due to conditioning on the absorption event at the end of limb 1) are then found using Bayes theorem
\begin{align*}
\hat{p}_{v\to w} &= \mathbb{P}[v \to w\ | \text{ absorbed at } (1,L_1)\text{ starting from }v ]\\
&= \frac{\mathbb{P}[v \to w\ \cap \text{ absorbed at } (1,L_1)\text{ starting from }v ]}{\mathbb{P}[\text{ absorbed at } (1,L_1)\text{ starting from }v ]}
\\
&= \frac{\mathbb{P}[\text{ absorbed at } (1,L_1)\text{ starting from }v \ | v \to w\ ] \mathbb{P}[v \to w]}{\mathbb{P}[\text{ absorbed at } (1,L_1)\text{ starting from }v ]}
\\
&= \frac{\mathbb{P}[\text{ absorbed at } (1,L_1)\text{ starting from }w\ ] }{\mathbb{P}[\text{ absorbed at } (1,L_1)\text{ starting from }v\ ]}p_{vw}
\\
&= \frac{h_w}{h_v}p_{v\to w},
\end{align*} 
where $h_w$ and $h_v$ are the previously found conditional hitting probabilities starting at node $w$ and $v$, respectively. This allows us to derive the expected number of steps till absorption at a specific limb for the original random walker, defined as
\begin{subequations}
\begin{align}
\hat{k}_{i,j} &= \mathbb{E}[\ n \ |\ \text{ absorbed at } (1,L_1)\text{ starting from } (i,j)\ ],\\
\hat{k}_0 &= \mathbb{E}[ \ n \ |\ \text{ absorbed at } (1,L_1) \text{ starting from the origin}\ ],
\end{align} 
\end{subequations}
by considering the number of steps till absorption for the biased random walker. We use the observation that the number of steps till absorption starting from any of the other endpoints of the star graph, i.e.\ $(i,L_i)$ for $i\neq 1$, can be safely set to zero by noting that conditioning on reaching $(1,L_1)$ makes it impossible to reach these ends because $h_{i,L_i}=0$. This gives the following set of algebraic equations describing the $\hat{k}_{i,j}$
\begin{subequations}
\begin{align}
\hat{k}_{i,L_i} & = 0, \quad \text{for }1\leq i\leq M, \\
\hat{k}_0 &= 1 + \sum_{i=1}^M \hat{p}_{0\to (i,1)} \hat{k}_{i,1},\\
\hat{k}_{i,j} &=1+ \hat{p}_{(i,j) \to (i,j-1)} \hat{k}_{i,j-1} + \hat{p}_{(i,j)\to(i,j+1)}\hat{k}_{i,j+1}, \quad \text{for }1\leq i\leq M\text{ and }1\leq j \leq L_i-1.
\end{align}
\end{subequations}	
Using \ref{eq:absorption_prob} to find the explicit expressions for the transition probabilities of the biased random walker we find the conditional absorption times
\begin{subequations}
\begin{align}
\hat{k}_0 &= \frac{1}{3}\left(L_1^2 + 2 h(\vec{L}) a(\vec{L})\right),\\
\hat{k}_{1,j} &= \frac{h(\vec{L})(L_1-j)}{h(\vec{L})(L_1-j) + jM L_1}\left(\hat{k}_0 + \frac{1}{3}j(2L_1-j) + j(L_1+j)\frac{M L_1}{3h(\vec{L})} \right), \quad \text{for }1\leq j \leq L_1,\\
\hat{k}_{i,j} &= \hat{k}_0 + \frac{1}{3}j(2 L_i - j), \quad \text{for }i \neq 1\text{ and }1\leq j \leq L_i-1.
\end{align} 
\end{subequations}

Note that these results are consistent with the unconditional mean absorption time results derived earlier, e.g.\
\begin{equation}
\sum_{i=1}^M \frac{h(\vec{L})}{M L_i} \times \frac{1}{3}\left(L_i^2 + 2 h(\vec{L}) a(\vec{L})\right) = h(\vec{L}) a(\vec{L}) = k_0.
\end{equation}

\paragraph{Second moment of absorption time} First we consider the problem of finding the expected square of the number of steps till absorption at any of the limb ends, defined as
\begin{subequations}
\begin{align}
v_{i,j} &= \mathbb{E}[\ n^2 \ |\ \text{ starting from } (i,j)\ ],\\
v_0 &= \mathbb{E}[ \ n^2 \ |\ \text{ starting from the origin}\ ],
\end{align} 
\end{subequations}
where again we identify $v_{i,0}\equiv v_0$ for all $1\leq i\leq M$. Similar to the previous sections we use conditioning on the next move of the walker to derive the set of algebraic equations describing the $v_{i,j}$
\begin{subequations}
\begin{align}
v_{i,L_i} & = 0, &\text{for }1\leq i\leq M, \\
v_0       &= 1 + \sum_{i=1}^M \frac{1}{M}v_{i,1} + 2 \sum_{i=1}^M\frac{1}{M}k_{i,1}, \\
v_{i,j}   &= 1 + \frac{1}{2} v_{i,j-1} + \frac{1}{2} v_{i,j+1} + k_{i,j-1} + k_{i,j+1}, & \text{for }1\leq i\leq M\text{ and }1\leq j \leq L_i-1.
\end{align}
\end{subequations}	
Using the results from \ref{eq:mean_star_graph} we then rewrite this system as
\begin{subequations}
\begin{align}
v_{i,L_i} & = 0, &\text{for }1\leq i\leq M, \\
v_0       &= 2k_0 - 1 + \sum_{i=1}^M \frac{1}{M}v_{i,1}, \\
v_{i,j}   &= 2k_{i,j}-1 + \frac{1}{2} v_{i,j-1} + \frac{1}{2}v_{i,j+1}, & \text{for }1\leq i\leq M\text{ and }1\leq j \leq L_i-1.
\end{align}
\end{subequations}	

This system is solved by 
\begin{subequations}\label{eq:2nd_moment_star_graph}
\begin{align}
v_0 &= \frac{4}{3}k_0^2 - \frac{2}{3}k_0 + \frac{1}{3} h(\vec{L}) a(\vec{L}^3),\\
v_{i,j} &= \left(1-\frac{j}{L_i}\right)\left(v_0 - \frac{1}{3} j \left(2j k_0 + (2 + j^2 - 4k_0)L_i - j L_i^2 - L_i^3 \right)\right),  \quad \text{for }1\leq j \leq L_i,
\end{align} 
\end{subequations}
where $a(\vec{L}^3)=M^{-1}\sum_m L_m^3$ is the arithmetic mean of the cubed limb lengths $L_m^3$.

\subsubsection{Continuous random walk model}

Here we detail how to get the first passage properties for a stargraph with limbds of non-integer length, representing continuous diffusion on the star graph. First, ;et $\ell_i = L_i \Delta x$, be the lengths of each of the limbs. Then rescale the first and second moments of first passage times by $\Delta t$ and $\Delta t^2$, respectively, where $\Delta t$ is an arbitrary time step. Then for the absorption probabilities as well as the moments of first passage times take the following limit $\Delta x \to 0$ while $\Delta t = \Delta x^2/2D$, where $D$ is the diffusion coefficient of the continuous diffusive process. The first passage properties on a star graph become

\begin{subequations}\label{eq:absorption_prob_CTRW}
\begin{align}
    h_{i}(x_j) &= \begin{cases} h_0(1-x_j/\ell_i), &\qquad \text{if } i\neq 1\\
        h_0(1-x_j/\ell_1) + x_j/\ell_1, &\qquad \text{if } i=1
    \end{cases},\\
    h_0 &= \frac{h(\vec{\ell})}{M\ell_1}
\end{align} 
\end{subequations}

\begin{subequations}\label{eq:mean_star_graph_CTRW}
\begin{align}
\mathbb{E}(\tau) &= \lim_{\substack{\Delta t\to 0, \Delta x\to 0\\ \Delta t = \Delta x^2/2D}}k_0\Delta t&= \frac{h(\vec{\ell}) a(\vec{\ell})}{2D},\\
\mathbb{E}(\tau) &= \lim_{\substack{\Delta t\to 0, \Delta x\to 0\\ \Delta t = \Delta x^2/2D}}k_{i,j} \Delta t &= \left(1-\frac{x_j}{\ell_i}\right)\left(k_0 + \frac{x_j \ell_i}{2D}\right), \quad \text{for }0 \leq x_j \leq \ell_i,
\end{align} 
\end{subequations}

\begin{subequations}\label{eq:2nd_moment_star_graph_CTRW}
\begin{align}
v_0(\Delta t)^2 &= \frac{1}{3D^2}\left(h(\vec{\ell})a(\vec{\ell})\right)^2 + \frac{1}{12D^2} h(\vec{\ell}) a(\vec{\ell}^{\ 3}),\\
v_{i,j}(\Delta t)^2 &= \left(1-\frac{x_j}{\ell_i}\right)\left(v_0 - \frac{1}{3} x_j \left(2 x_j k_0 + (x_j^2 - 4k_0)\ell_i - x_j \ell_i^2 - \ell_i^3 \right)\right),  \quad \text{for }0\leq x_j \leq \ell_i,
\end{align} 
\end{subequations}

\subsection{Deriving moments of off-network first passage times}
\label{appendix:off-network-moments}

In this section we briefly introduce how to construct and solve a hierarchy of equations to provide first passage time moments for a planar diffusive transport process. This derivation is an adaptation of the derivation as seen in \cite{Redner2001_Book}. Firstly let us consider the $k$-th moments of a general first passage time problem. Let $f(t)$ and $S(t)$ denote a general first passage time distribution and survival probability at time $t$. The $k$-th moment of the first passage time is defined as follows
\begin{equation}\label{Aeq:kthmoment}
\mathbb{E} \left( t^k \right) = \int_{0}^{\infty} t^k f(t) \mathrm{d}t.
\end{equation}
Noting that the CDF of the first passage time is given by $1-S(t)$, we can differentiate the CDF and we find that $f(t) = - \mathrm{d}S(t)/\mathrm{d}t$. Substituting this expression for $f(t)$ into Eq.~\eqref{Aeq:kthmoment} and integrating by parts yields
\begin{equation}\label{Aeq:kthmoment_2}
\mathbb{E} \left( t^k \right) = k \int_{0}^{\infty} t^{k-1} S(t) \mathrm{d}t.
\end{equation}
Let $p(\mathbf{r},t)$ define the probability density of finding a particle at position vector $\mathbf{r}$ at time $t$. Then the survival probability that the particle has not left an enclosed domain $\Omega$ is given by $S(t) = \int_{\Omega} p(\mathbf{r},t) \mathrm{d}\mathbf{r}$. Substituting this expression for $S(t)$ into Eq.~\eqref{Aeq:kthmoment_2} we arrive at the following formula for the $k$-th moment of the first passage time.
\begin{equation}\label{Aeq:kthmoment_3}
\mathbb{E} \left( t^k \right) = k \int_{\Omega} C_{k-1} \left( \mathbf{r}\right) \mathrm{d} \mathbf{r},
\end{equation}
where 
\begin{equation}
C_{k} \left( \mathbf{r}\right) = \int_{0}^{\infty} t^k p(\mathbf{r},t) \mathrm{d}t.
\end{equation}
In order to calculate the first passage moments we now derive a hierarchy of equations to solve for $C_{k} \left( \mathbf{r}\right)$.

The isotropic diffusion equation in $d$-dimensional Euclidean space is given by
\begin{equation}\label{Aeq:diffusionequation}
\dfrac{\partial p(\mathbf{r},t)}{\partial t} = D \nabla^2 p(\mathbf{r},t),
\end{equation}
where $\mathbf{r} = \left(x_1,\ldots,x_d\right) \in \mathbb{R}^d$ and $\nabla^2 = \left( \partial^2 / \partial x_1^2,\ldots,\partial^2 / \partial x_d^2\right)$. Multiplying the right-hand side of Eq.~\eqref{Aeq:diffusionequation} by $t^k$ and integrating over time from zero to infinity yields $D \nabla^2 C_{k}(\mathbf{r})$. Repeating this transformation to the left-hand side and using integration by parts yields $\left[ t^k p(\mathbf{r},t) \right]_0^{\infty} - k \int_{0}^{\infty} t^{k-1} p(\mathbf{r},t) \mathrm{d}t$. Thus noting that unless $k=0$ the integrated part vanishes we write write down the following hierarchy of equations
\begin{subequations}\label{Aeq:hierarchy}
\begin{eqnarray}
D \nabla^2 C_0 \left( \mathbf{r}\right) &=& - \delta \left(\mathbf{r}-\mathbf{r}_0 \right) \label{Aeq:C0}\\
D \nabla^2 C_1 \left( \mathbf{r}\right) &=& - C_0 \left( \mathbf{r}\right) \\
&\vdots& \\
D \nabla^2 C_k \left( \mathbf{r}\right) &=& - k C_{k-1} \left( \mathbf{r}\right), 
\end{eqnarray}
\end{subequations}
where $\mathbf{r}_0$ is the initial position of the particle. In fact, we can combine Eq.~\eqref{Aeq:kthmoment_3} and Eqs.~\eqref{Aeq:hierarchy} to derive a set of hierarchical equations that solves for the moments directly. Let $\rho^{(k)}(\mathbf{r}_0) =\mathbb{E} \left( t^k \right)$ where we note that first passage moments will depend upon the initial position of the particle. Taking the Laplacian with respect to the initial coordinate $\mathbf{r}_0$ and using Eqs.~\eqref{Aeq:hierarchy} yields
\begin{subequations}\label{Aeq:hierarchy_direct}
\begin{eqnarray}
D \nabla^2_{\mathbf{r}_0} \rho^{(1)} \left( \mathbf{r}_0\right) &=& - 1 \label{Aeq:rho_1}\\
D \nabla^2_{\mathbf{r}_0} \rho^{(2)} \left( \mathbf{r}_0\right) &=& - 2 \rho^{(1)} \left( \mathbf{r}_0\right) \\
&\vdots& \\
D \nabla^2_{\mathbf{r}_0} \rho^{(k)} \left( \mathbf{r}_0\right) &=& - k \rho^{(k-1)}\left( \mathbf{r}_0\right),
\end{eqnarray}
\end{subequations}
\subsubsection{First passage moments within a $d$-dimensional sphere}
In this section we derive the first and second moments of the time taken for a diffusing particle within a $d$-dimensional sphere of radius $R$ to first hit the boundary as a function of its initial position. We assume spherical symmetry (as is appropriate when the initial condition is the origin) and write the Laplacian in spherical coordinates and Eq.~\eqref{Aeq:rho_1} becomes the following ODE
\begin{equation}\label{eqA:ODE_moment1}
\dfrac{D}{r_0^{d-1}} \dfrac{\mathrm{d}}{\mathrm{d}r_0} \left( r_0^{d-1} \dfrac{\mathrm{d}\rho^{(1)}(r_0)}{\mathrm{d}r_0}\right) = - 1
\end{equation}
For boundary conditions $\rho^{(1)}(R)=0$, and we assume that $\mathrm{d}\rho^{(1)}/\mathrm{d}r_0(0)=0$ to avoid singularities at the origin. Equation \eqref{eqA:ODE_moment1} has the solution
\begin{equation}\label{Aeq:FirstMoment}
\rho^{(1)} \left( r_0 \right) = \dfrac{R^2 - r_0^2}{2dD}.
\end{equation}
For the second moment we solve the equation
\begin{equation}
\dfrac{D}{r_0^{d-1}} \dfrac{\mathrm{d}}{\mathrm{d}r_0} \left( r_0^{d-1} \dfrac{\mathrm{d}\rho^{(2)}(r_0)}{\mathrm{d}r_0}\right) = - \dfrac{R^2 - r_0^2}{dD},
\end{equation}
with similar boundary conditions $\rho^{(2)}(R)=0$ and $\mathrm{d}\rho^{(2)}/\mathrm{d}r_0(0)=0$ which has solution
\begin{equation}\label{Aeq:SecondMoment}
\rho^{(2)} \left( r_0 \right) = \dfrac{R^2\left(R^2 - r_0^2\right)}{2d^2D^2} + \dfrac{r_0^4 - R^4}{4(d+2)dD^2}.
\end{equation}
\subsubsection{Coefficient of Variation}
In this section we provide the formula for the coefficient of variation for the first passage time for a diffusing particle in a sphere initially at the origin. The coefficient of variation is defined as the ratio of the standard deviation over the mean. Thus we can calculate the coefficient of variation $\mathcal{CV} = \left(\rho^{(2)}(0)-\rho^{(1)}(0)^2\right)^{1/2} / \rho^{(1)}(0)$ using Eqs.~\eqref{Aeq:FirstMoment} and \eqref{Aeq:SecondMoment}, such that
\begin{equation}
\mathcal{CV} = \sqrt{\dfrac{2}{2+d}}.
\end{equation}
Note that the coefficient of variation only depends on the dimension of the Euclidean space. Thus we can fit the networked coefficient of variation to produce an "effective dimension" for diffusion on a given network.
\subsubsection{Radially dependent diffusion}
In this section we derive the first two moments for the first passage time for a particle whose position evolves according to a diffusion equation with radially dependent diffusion coefficient,
\begin{equation}\label{Aeq:radialdiffusionequation}
\dfrac{\partial p(r,t)}{\partial t} = D r^{d_f-1} \dfrac{\partial}{\partial r} \left \lbrace r^{d_f-1-\Theta} \dfrac{\partial p(r,t)}{\partial r} \right \rbrace.
\end{equation}
The hierarchy of equations to solve for the first passage moments can be derived analogously and the equation for the first moment $\rho^{(1)}(r_0)$ is
\begin{equation}
D r_0^{d_f-1} \dfrac{\mathrm{d}}{\mathrm{d} r_0} \left( r_0^{d_f-1-\Theta} \dfrac{\mathrm{d} \rho^{(1)}(r_0)}{\mathrm{d} r_0} \right) = -1,
\end{equation}
with boundary conditions $\rho^{(1)}(R)=0$ and $\mathrm{d}\rho^{(1)}/\mathrm{d}r_0(0)=0$. The solution is given by
\begin{equation}\label{Aeq:rho1_radial}
\rho^{(1)}(r_0) = \dfrac{R^{2+\Theta}-r_0^{2+\Theta}}{\left( 2+\Theta \right)d_f D}.
\end{equation}
Similarly, for the second we have the equation
\begin{equation}
D r_0^{d_f-1} \dfrac{\mathrm{d}}{\mathrm{d} r_0} \left( r_0^{d_f-1-\Theta} \dfrac{\mathrm{d} \rho^{(2)}(r_0)}{\mathrm{d} r_0} \right) = -2\rho^{(1)}(r_0),
\end{equation}
with boundary conditions $\rho^{(2)}(R)=0$ and $\mathrm{d}\rho^{(2)}/\mathrm{d}r_0(0)=0$. The solution here is given by
\begin{equation}\label{Aeq:rho2_radial}
\rho^{(2)}(r_0) = \dfrac{r_0^{4+2\Theta}}{\left( 2+\Theta\right)^2 \left( 2+\Theta+d_f\right)d_f D^2} - 2\dfrac{R^{2+\Theta}r_0^{2+\Theta}}{\left( 2+\Theta\right)^2 d_f^2 D^2} + \dfrac{R^{4+2\Theta}\left( 4+2\Theta+d_f\right)}{\left( 2+\Theta\right)^2\left( 2+\Theta+d_f\right)d_f^2D^2}.
\end{equation}
The coefficient of variation can be calculated from Eqs.~\eqref{Aeq:rho1_radial} and \eqref{Aeq:rho2_radial} to provide the formula
\begin{equation}
\mathcal{CV} = \sqrt{\dfrac{2+\Theta}{2+\Theta+d_f}}.
\end{equation}
\subsubsection{Skewness and Kurtosis}
In this section we present the skewness and kurtosis of the first passage time for a particle in the sphere initially at the origin undergoing radially dependent diffusion. The standardised moments of the first passage time distribution are given by the ratio of the $k$-th moment and the standard deviation to the $k$-th power, i.e. $\rho^{(k)}\left( r_0 \right) / \left( \rho^{(2)}\left( r_0 \right) - \rho^{(1)}\left( r_0 \right)^2\right)^{k/2}$. The skewness and kurtosis are given by the third and fourth standardised moments, respectively. Solving the hierarchy of equations \eqref{Aeq:hierarchy_direct} with radially dependent diffusion coefficient reveals that the skewness, $\mathcal{SK}$, is given by
\begin{equation}
\mathcal{SK} = \dfrac{4\sqrt{\left(2+\Theta\right)\left(2+\Theta+d_f\right)}}{4+d_f+2\Theta},
\end{equation}
and the kurtosis, $\mathcal{KU}$, is given by
\begin{equation}
\mathcal{KU} = 9 - \dfrac{6d_f^2}{\left( 4+d_f+2\Theta\right) \left( 6+d_f+3\Theta\right)}.
\end{equation}

\end{document}